\documentclass[3p,times,book,onecolumn]{elsarticle}

\usepackage{titlesec}
\titleclass{\subsubsubsection}{straight}[\subsection]

\newcounter{subsubsubsection}[subsubsection]
\renewcommand\thesubsubsubsection{\thesubsubsection.\arabic{subsubsubsection}}

\titleformat{\subsubsubsection}
  {\normalfont\normalsize\itshape}{\thesubsubsubsection}{1em}{}
\titlespacing*{\subsubsubsection}
{0pt}{3.25ex plus 1ex minus .2ex}{.5ex plus .2ex}

\makeatletter
\renewcommand\paragraph{\@startsection{paragraph}{5}{\z@}%
  {3.25ex \@plus1ex \@minus.2ex}%
  {-1em}%
  {\normalfont\normalsize\bfseries}}
\renewcommand\subparagraph{\@startsection{subparagraph}{6}{\parindent}%
  {3.25ex \@plus1ex \@minus .2ex}%
  {-1em}%
  {\normalfont\normalsize\bfseries}}

\titleclass{\subsubsubsubsection}{straight}[\subsection]

\newcounter{subsubsubsubsection}[subsubsubsection]
\renewcommand\thesubsubsubsubsection{\thesubsubsubsection.\arabic{subsubsubsubsection}}

\titleformat{\subsubsubsubsection}
  {\normalfont\normalsize\itshape}{\thesubsubsubsubsection}{1em}{}
\titlespacing*{\subsubsubsubsection}
{0pt}{3.25ex plus 0ex minus .2ex}{.5ex plus .2ex}

\makeatletter
\renewcommand\paragraph{\@startsection{paragraph}{6}{\z@}%
  {3.25ex \@plus1ex \@minus.2ex}%
  {-1em}%
  {\normalfont\normalsize\bfseries}}
\renewcommand\subparagraph{\@startsection{subparagraph}{7}{\parindent}%
  {3.25ex \@plus1ex \@minus .2ex}%
  {-1em}%
  {\normalfont\normalsize\bfseries}}
\def\toclevel@subsubsubsection{4}
\def\toclevel@subsubsubsubsection{5}
\def\toclevel@paragraph{6}
\def\toclevel@paragraph{7}
\def\l@subsubsubsection{\@dottedtocline{4}{7em}{4em}}
\def\l@subsubsubsubsection{\@dottedtocline{5}{10em}{5em}}
\def\l@paragraph{\@dottedtocline{6}{14em}{6em}}
\def\l@subparagraph{\@dottedtocline{7}{17em}{7em}}
\makeatother

\usepackage[utf8]{inputenc}
\usepackage[english]{babel}

\usepackage{graphicx,epsfig,amsmath,color}

\usepackage{multirow}
\usepackage{cuted}
\usepackage{siunitx}
\usepackage{mhchem}
\usepackage{caption}
\usepackage{subcaption}
\usepackage{mathtools}
\usepackage{amsmath}
\usepackage{hyperref}
\usepackage{booktabs}
\usepackage{float}
\usepackage{gensymb}

\usepackage{lineno}
\definecolor{red}{rgb}{1,0,0}
\definecolor{blue}{rgb}{0,0,1}

\makeatletter
\def\ps@pprintTitle{%
  \let\@oddhead\@empty
  \let\@evenhead\@empty
  \def\@oddfoot{\reset@font\hfil\thepage\hfil}
  \let\@evenfoot\@oddfoot
}
\makeatother

\begin{document}

\begin{frontmatter}
\begin{titlepage}

\title{\textbf{The European Spallation Source neutrino Super Beam} \\
       \vspace{0.5cm} \large \it{White Paper to be submitted to the Snowmass 2021 \\ 
       USA Particle Physics Community Planning Exercise}}

\newcommand{\authorlist}{

\author[cern,uu]{A.~Alekou}
\author[iphc]{E.~Baussan}
\author[ess]{N.~Blaskovic Kraljevic}
\author[kth,okc]{M.~Blennow}
\author[unisof]{M.~Bogomilov}
\author[iphc]{E.~Bouquerel}
\author[ulund1]{A.~Burgman}
\author[uu2]{C.J.~Carlile}
\author[ulund1]{J.~Cederkall}
\author[ulund1]{P.~Christiansen}
\author[ulund2,ess]{M.~Collins}
\author[infn2]{E.~Cristaldo Morales}
\author[agh]{P.~Cupial}
\author[iphc]{L.~D'Alessi}
\author[ess]{H.~Danared}
\author[iphc]{J.~P.~A.~M.~de~Andr\'{e}}
\author[cern]{J.P.~Delahaye}
\author[iphc]{M.~Dracos\corref{contact}}
\author[cern]{I.~Efthymiopoulos}
\author[uu]{T.~Ekel\"{o}f\corref{contact}}
\author[ess]{M.~Eshraqi}
\author[ncsr]{G.~Fanourakis}
\author[uam]{E.~Fernandez-Martinez}
\author[ess]{B.~Folsom}
\author[ess]{N.~Gazis}
\author[ncsr]{Th.~Geralis}
\author[rbi]{M.~Ghosh}
\author[cu]{G.~Gokbulut}
\author[rbi]{L.~Halić}
\author[cu]{A.~Kayis Topaksu}
\author[ess]{B.~Kildetoft}
\author[rbi]{B.~Kliček}
\author[agh]{M.~Koziol}
\author[rbi]{K.~Krhač}
\author[agh]{L.~Lacny}
\author[ess]{M.~Lindroos}
\author[infn1]{M.~Mezzetto}
\author[cu]{M.~Oglakci}
\author[kth,okc]{T.~Ohlsson}
\author[uu]{M.~Olveg\r{a}rd}
\author[uam]{T.~Ota}
\author[ulund1]{J.~Park\fnref{ibs}}
\author[ess]{D.~Patrzalek}
\author[unisof]{G.~Petkov}
\author[iphc]{P.~Poussot}
\author[ess]{R.~Johansson}
\author[uam,IJC]{S.~Rosauro-Alcaraz}
\author[agh]{B.~Szybiński}
\author[agh]{J.~Snamina}
\author[ncsr]{G.~Stavropoulos}
\author[rbi]{M.~Stipčević}
\author[infn2]{F.~Terranova}
\author[iphc]{J.~Thomas}
\author[uhh]{T.~Tolba\corref{contact}}
\author[ess]{E.~Trachanas}
\author[unisof]{R.~Tsenov}
\author[unisof]{G.~Vankova-Kirilova}
\author[csns]{N.~Vassilopoulos} 
\author[cern]{E.~Wildner}
\author[iphc]{J.~Wurtz}
\author[ncsr]{O.~Zormpa}
\author[uu]{Y.~Zou}

\address[agh]{AGH University of Science and Technology, al. Mickiewicza 30, 30-059 Krakow, Poland}

\address[cu]{University of Cukurova, Faculty of Science and Letters, Department of Physics, 01330 Adana, Turkey}

\address[ncsr]{Institute of Nuclear and Particle Physics, NCSR Demokritos, Neapoleos 27, 15341 Agia Paraskevi, Greece}


\address[csns]{Spallation Neutron Science Center, Dongguan 523803, China} 


\address[cern]{CERN, 1211 Geneva 23, Switzerland}

\address[uhh]{Institute for Experimental Physics, Hamburg University, 22761 Hamburg, Germany}

\address[ulund1]{Department of Physics, Lund University, P.O Box 118, 221 00 Lund, Sweden}

\address[ulund2]{Faculty of Engineering, Lund University, P.O Box 118, 221 00 Lund, Sweden}

\address[ess]{European Spallation Source, Box 176, SE-221 00 Lund, Sweden}

\address[uam]{Departamento de Fisica Teorica and Instituto de Fisica Teorica, IFT-UAM/CSIC, Universidad Autonoma de Madrid, Cantoblanco, 28049, Madrid, Spain}

\address[infn2]{University of Milano-Bicocca and INFN sez. di Milano-Bicocca, Milano, Italy}

\address[infn1]{INFN sez. di Padova, Padova, Italy}

\address[uniri]{University of Rijeka, Department of Physics, 51000 Rijeka, Croatia}

\address[unisof]{Sofia University St. Kliment Ohridski, Faculty of Physics, 1164 Sofia, Bulgaria}

\address[iphc]{IPHC, Universit\'{e} de Strasbourg, CNRS/IN2P3, Strasbourg, France}

\address[kth]{Department of Physics, School of Engineering Sciences, KTH Royal Institute of Technology, Roslagstullsbacken 21, 106 91 Stockholm, Sweden}

\address[okc]{The Oskar Klein Centre, AlbaNova University Center, Roslagstullsbacken 21, 106 91 Stockholm, Sweden}

\address[uu]{Uppsala University, P.O. Box 256, 751 05 Uppsala, Sweden}

\address[uu2]{Department of Physics and Astronomy, FREIA, Uppsala University, Box 516, 751 20 Uppsala, Sweden}

\address[rbi]{Center of Excellence for Advanced Materials and Sensing Devices, Ruđer Bo\v{s}kovi\'c Institute, 10000 Zagreb, Croatia}

\address[IJC]{P\^ole Th\'eorie, Laboratoire de Physique des 2 Infinis Ir\'ene Joliot Curie (UMR 9012) CNRS/IN2P3, 15 rue Georges Clemenceau, 91400 Orsay, France}

\cortext[contact]{Corresponding authors: marcos.dracos@in2p3.fr, tord.ekelof@physics.uu.se, tamer.tolba@uni-hamburg.de}

\fntext[ibs]{Now at The center for Exotic Nuclear Studies, Institute for Basic Science, 34126 Daejeon, Korea}

}

\authorlist

\begin{abstract}
\pagenumbering{roman}
In this Snowmass 2021 white paper, we summarise the Conceptual Design of the European Spallation Source neutrino Super Beam (ESS$\nu$SB) experiment and its synergies with the possible future muon based facilities, e.g. a Low Energy nuSTORM and the Muon Collider. The ESS$\nu$SB will benefit from the high power, \SI{5}{\mega\watt}, of the European Spallation Source (ESS) LINAC in Lund-Sweden to produce the world’s most intense neutrino beam, enabling measurements to be made at the second oscillation maximum. Assuming a ten-year exposure, physics simulations show that the CP-invariance violation can be established with a significance of 5$\sigma$ over more than \SI{70}{\percent} of all values of $\delta_{CP}$ and with an error in the measurement of the $\delta_{CP}$ angle of less than 8$\degree$ for all values of  $\delta_{CP}$.\\

However, several technological and physics challenges must be further studied before achieving a final Technical Design. Measuring at the 2nd oscillation maximum necessitates a very intense neutrino beam with the appropriate energy. For this, the ESS proton beam LINAC, which is designed to produce the world's most intense neutron beam, will need to be upgraded to \SI{10}{\mega\watt} power, \SI{2.5}{\giga\electronvolt} energy and \SI{28}{\hertz} beam pulse repetition rate. An accumulator ring will be required for the compression of the ESS LINAC beam pulse from \SI{2.86}{\milli\second} to \SI{1.3}{\micro\second}. A high power target station facility will be needed to produce a well-focused intense (super) $\mu$-neutrino beam. The physics performance of that neutrino Super Beam in conjunction with a megaton underground Water Cherenkov neutrino far detector installed at a distance of either \SI{360}{\kilo\meter} or \SI{540}{\kilo\meter} from the ESS, the baseline, has been evaluated.

\end{abstract}


\end{titlepage}

\end{frontmatter}

\tableofcontents
 \pagebreak

\setcounter{footnote}{0}

\section{Introduction} \label{introduction}
\pagenumbering{arabic}

In the standard three flavour scenario, the phenomenon of neutrino oscillation can be described by three mixing angles: $\theta_{12}$, $\theta_{13}$ and $\theta_{23}$, two mass squared differences $\Delta$m$_{21}^{2}$ (= m$_{2}^{2}$ - m$_{1}^{2}$) and $\Delta$m$_{31}^{2}$ (= m$_{3}^{2}$ - m$_{1}^{2}$) and one Dirac type phase $\delta_{CP}$. During the past few decades, some of these parameters were measured with good precision. At the moment, the unknown parameters are: (i) the mass hierarchy of the neutrinos, which can be either normal, i.e. $\Delta$m$_{31}^{2}$ $>$ 0, or inverted, i.e. $\Delta$m$_{31}^{2}$ $<$ 0, (ii) the octant of the mixing angle $\theta_{23}$, which can be either at the lower octant, i.e. $\theta_{23}$ $<$ 45$\degree$, or at the higher octant, i.e. $\theta_{23}$ $>$ 45$\degree$ and (iii) the Dirac violating phase, $\delta_{CP}$.\\ 

\begin{figure}[H]
\centering
\includegraphics[width=7.cm]{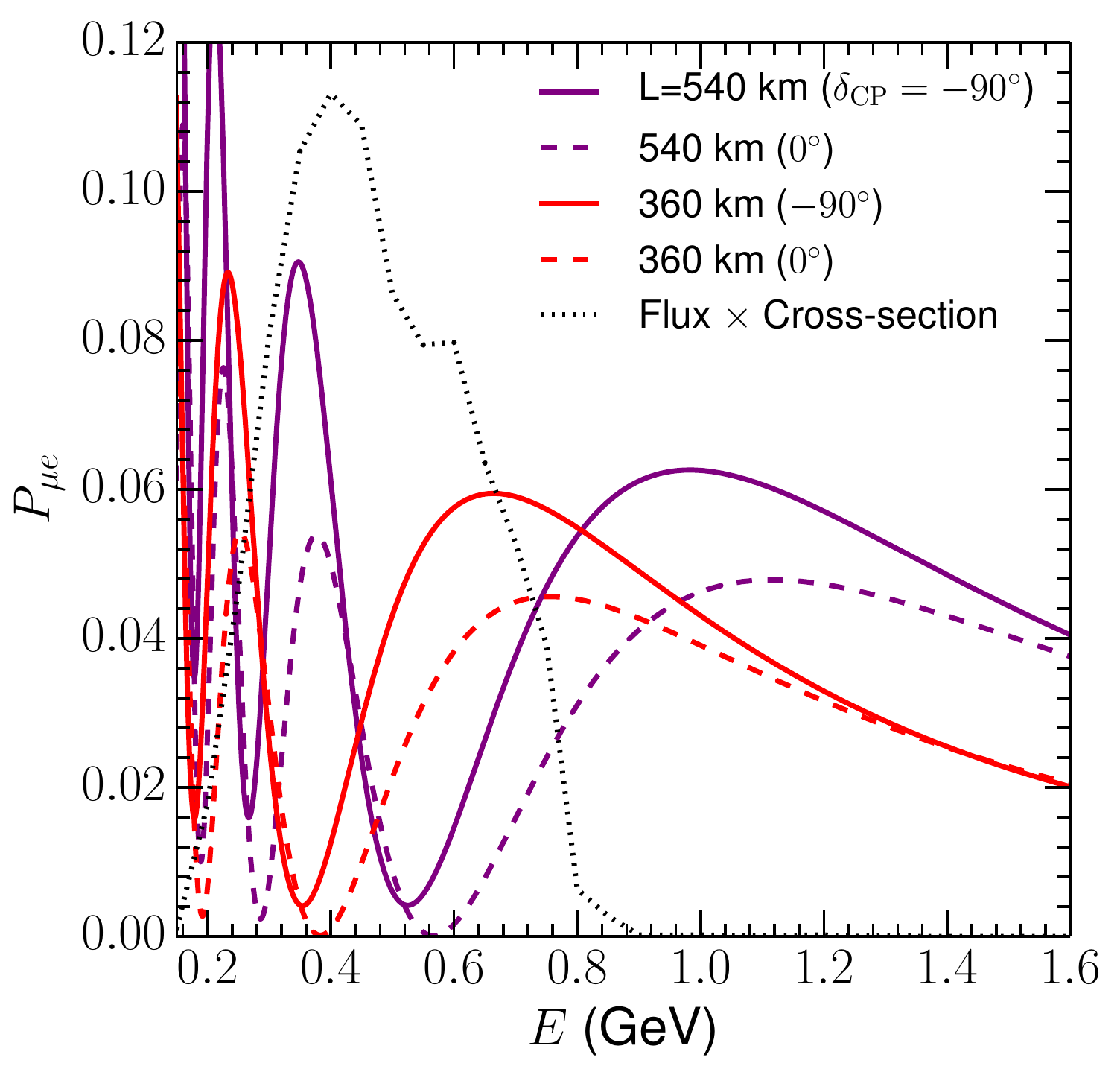}
\includegraphics[width=7.cm]{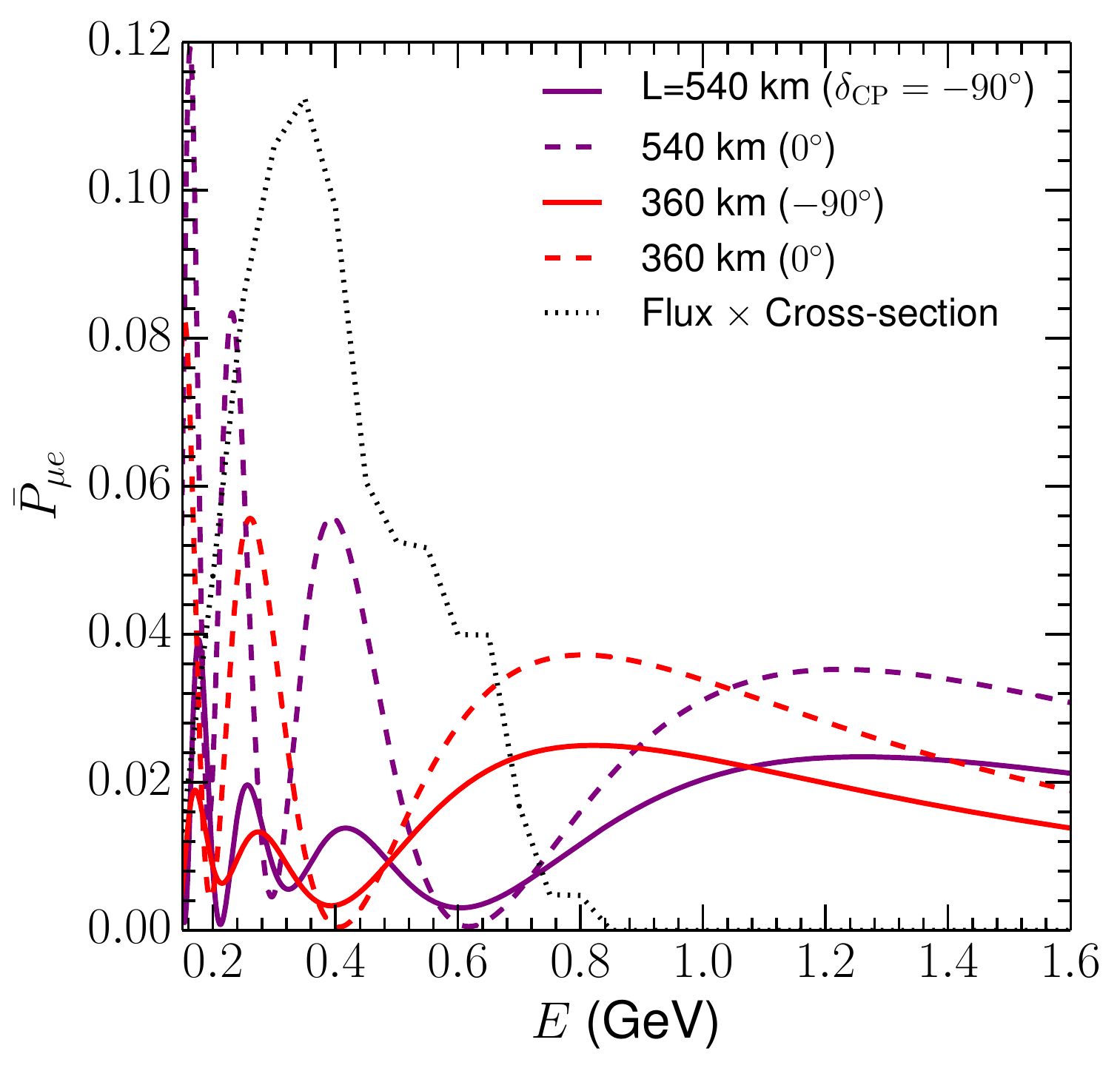}
\caption{Oscillation probabilities for the \SI{360}{km} (red curves) and \SI{540}{km} (purple curves) baselines as a function of the energy for neutrinos (left panel) and antineutrinos (right panel). The solid (dotted) lines are for $\delta = -90^\circ$ ($\delta =0$). The grey dotted lines show the convolution of the signal component of the neutrino flux with the detection cross section. Thus, they serve as a guide of what energies of the oscillation probability would be well-sampled by the ESS$\nu$SB setup.}
\label{fig:probs}
\end{figure}

In the search for the CP-violation in the leptonic sector, crucial information has been obtained from reactor and accelerator experiments \cite{An:2012eh,Abe:2011sj}. The discovery and measurement of the third neutrino mixing angle, $\theta_{13}$, with a value $\sim$~\SI{9}{\degree}, corresponding to $\sin^{2}$2$\theta_{13}$~$\sim$~0.095, confirmed the possibility of discovering and measuring a non-zero value of the Dirac leptonic CP violating angle, $\delta_{CP}$. Before this measurement, a significantly smaller value of $\theta_{13}$ was assumed, with a range of values of $\sin^{2}$2$\theta_{13}$~$\sim$~ 0.01 and 0.09, with 0.04 as a standard value. In the light of this new finding, the sensitivity to CP violation observation and measurement, with precision, of $\delta_{CP}$ has shown a strong enhancement at the second oscillation maximum compared to that at the first oscillation maximum~\cite{Nunokawa:2007qh, Coloma:2011pg, Parke:2013pna}. This can be seen in Figure~\ref{fig:probs} where the change in the probability, upon changing the values of $\delta_{CP}$, is much more significant at the second peak maximum compared to the first. Moreover, by placing the far detector at the second oscillation maximum, the experiment is significantly less affected by, and hence more robust against, systematic uncertainties. This is particularly important since the improvement of the present systematic errors is known to be very hard. However, placing the far detector at the 2nd oscillation maximum implies the need to use very high intensity "super" neutrino beams to compensate for the longer baseline, hence lower statistics.

The long-baseline experiments which are currently running to measure these unknowns are T2K \cite{Abe:2020vii} in Japan and NO$\nu$A \cite{Acero:2019ksn} in the USA. Regarding the true hierarchy of the neutrino mass, the results of both T2K and NO$\nu$A favour normal hierarchy over inverted hierarchy. Regarding the true nature of the octant of $\theta_{23}$, both these experiments support a higher octant, however the maximal value, i.e. $\theta_{23}$ = 45$\degree$, is also allowed within 1$\sigma$. Regarding the value of $\delta_{CP}$, there is a mismatch between T2K and NO$\nu$A. Considering the branch for $\delta_{CP}$ as -180$\degree$ $\leq$ $\delta_{CP}$ $\leq$ 180$\degree$, T2K supports the best-fit value of $\delta_{CP}$ around -90$\degree$, i.e. maximal CP violation, and the best-fit value measured by NO$\nu$A is around 0$\degree$, i.e. CP conservation. However, it is important to note that both the values of $\delta_{CP}$ = 0$\degree$ and -90$\degree$ are allowed at 3$\sigma$ and it requires more data to establish the true value of $\delta_{CP}$. It is believed that these two experiments will give a hint towards the true nature of these unknown oscillation parameters, while the future generation of the long-baseline experiments will establish these facts with a significant confidence level.\\

The European Spallation Source neutrino Super-Beam (ESS$\nu$SB) is a proposed accelerator-based long-baseline neutrino experiment in Sweden \cite{Baussan:2013zcy}. In this project, a high intensity neutrino beam will be produced at the upgraded ESS facility in Lund. These neutrinos will be detected at a distance of \SI{360}{\kilo\meter}, alternatively \SI{540}{\kilo\meter}, from the ESS site. The primary goal of this experiment is to measure the leptonic CP phase, $\delta_{CP}$, by probing the second oscillation maximum. As the variation of neutrino oscillation probability with respect to $\delta_{CP}$ is much higher in the second oscillation maximum, as compared to the first oscillation maximum \cite{Nunokawa:2007qh,Coloma:2011pg, Parke:2013pna}, ESS$\nu$SB, as a second generation super-beam experiment has the potential of measuring $\delta_{CP}$ with unprecedented precision compared to the first generation long-baseline experiments. 

\section{Instrumental Implementation}
\label{instrument}

\begin{figure}
    \centering
    \includegraphics[width=0.70\linewidth]{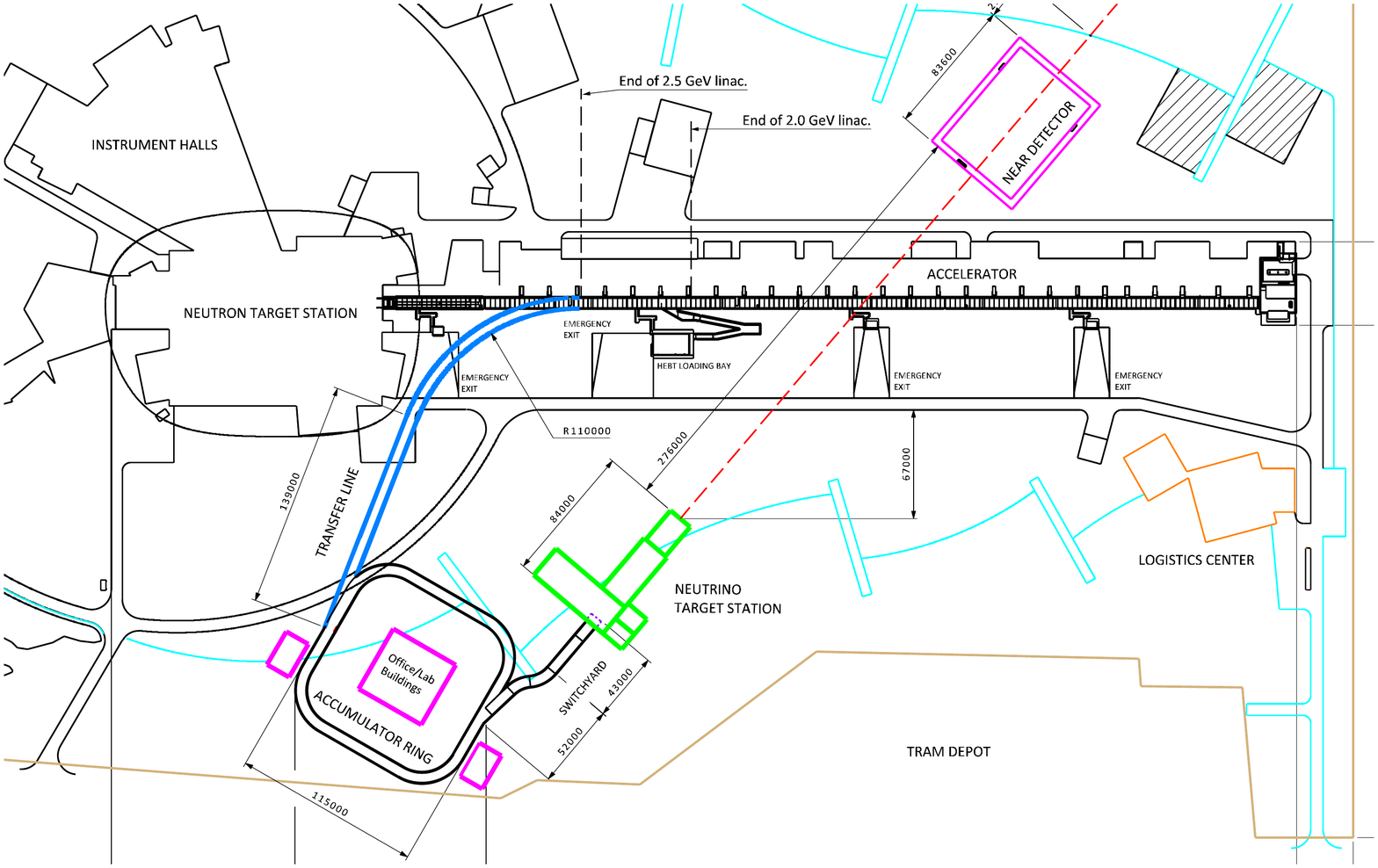}
    \caption{Schematic layout of the required ESS upgrades. The ESS linear accelerator is shown in black, the transfer line from the LINAC to the accumulator ring is shown in blue, the accumulator ring and the switchyard are shown in black, the target station is shown in green, and the near detector site is shown in magenta. (Units in mm)}
    \label{fig:ess-upgrade}
\end{figure}

To be able to produce an intense neutrino beam, concurrently with the intense neutron beam of the ESS, it is necessary to apply a number of upgrades to the ESS facility. The proposed upgrades are schematically presented in Figure~\ref{fig:ess-upgrade}. The pulse frequency of the ESS LINAC (the proton driver) must be increased from \SI{14}{\hertz} to \SI{28}{\hertz} to obtain additional acceleration cycles that will be used for neutrino production, without affecting the neutron programme. Moreover, during neutrino cycles, H$^{-}$ ions instead of protons need to be accelerated in order to ease the filling of the accumulator ring. An accumulator ring will be built to shorten the ESS pulse to about \SI{1.2}{\micro\second}. A neutrino production target station, composed of four identical targets enveloped by four magnetic focusing devices (horns), will be built. The horns will be used to charge select and focus the pions, and thus also the neutrinos resulting from their decay, toward the near and far detectors. A near detector will be used to monitor the neutrino beam and to measure neutrino interaction cross-sections, especially electron neutrino cross-sections, at a short baseline from the neutrino source, while the far detector will be used to detect the oscillated neutrino beam at the long baseline distance. In the following sections, these parts are discussed together with the physics potential of the experiment.

\subsection{\textbf{Proton Driver}}
\label{protondriver}

\begin{figure*}[hbt]
\begin{center}
\includegraphics[width=1\textwidth]{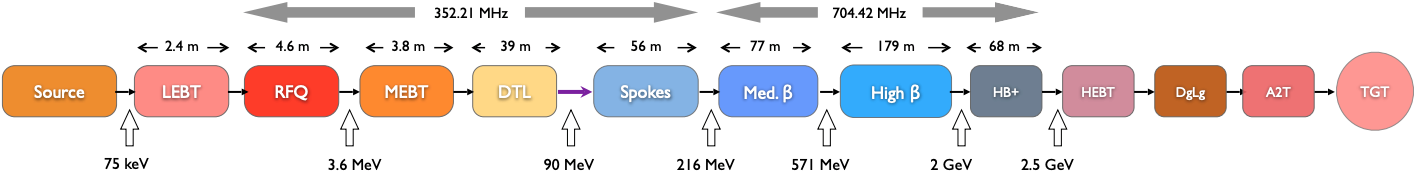}
\caption{\small Proton driver layout.}
\label{fig:protondriver}
\end{center}
\end{figure*}

The ESS LINAC, currently being constructed, will accelerate 14 proton pulses of \SI{2.86}{\milli\second} length per second. Each pulse contains about 10$^{15}$ protons, implying the production of a \SI{5}{\mega\watt} proton beam. Figure~\ref{fig:protondriver} shows the layout of the ESS proton driver. The proposed power increase of the LINAC from \SI{5}{\mega\watt} to \SI{10}{\mega\watt} will be realised by increasing the pulse frequency from \SI{14}{\hertz} to \SI{28}{\hertz}, adding 14 more H$^{-}$ pulses of the same length and number of particles, interleaved with the proton pulses. Each of the H$^{-}$ pulses will be chopped into four sub-pulses separated by gaps of $\sim$\SI{100}{\nano\second} length. The reason for accelerating H$^{-}$ ions is that they can be stripped of their electrons to inject protons in a manner that adiabatically increases the phase space density for an existing proton beam (otherwise limited by Liouville's theorem). The need for bunching is dictated by the rise and fall-time requirement of the extraction kicker magnets in the accumulator ring.\\ 

The total number of particles delivered to the accumulator ring will be $8.9\times 10^{14}$ per pulse cycle (macro-pulse), divided into four batches of $2.2 \times 10^{14}$. Each batch is stacked in the accumulator ring, compressing the \SI{2.86}{\milli\second} to four \SI{1.2}{\micro\second} pulses, which are subsequently extracted to the target. The acceleration of H$^{-}$ ions requires the addition of a H$^{-}$ ion source to the side of the LINAC proton source and a doubling of the front-end accelerator elements up to the point where the proton and H$^{-}$ beam-lines are merged. For the H$^-$ production, the Penning source~\cite{dudnikov:2012} at ISIS-RAL~\cite{isis} and the RF source at SNS~\cite{shishlo_2012,folsom:2021strp} were identified as the most promising ion sources to meet the ESS$\nu$SB requirements.

It is of primary interest that the beam exiting the LINAC is matched to the accumulator ring in order to have efficient injection and avoid losses. A LINAC-to-ring (L2R) transfer line has been designed to transport the \SI{2.5}{\giga\electronvolt} H$^-$ beam output from the upgraded high-$\beta$ line (HBL) at the end of the LINAC to the Accumulator Ring (AR) \cite{alekou:2020_D3.2}. At beam energies greater than ${\sim}$\SI{100}{\mega\electronvolt}, activation of machine components could become a concern if the loss values would exceed acceptable limits. The beam-loss will be kept below \SI{1}{\watt}/m, which ensures a maximum value of 1~mSv/h ambient dose rate due to activation at \SI{30}{\cm} from a surface of any given accelerator component, after 100~days of irradiation and \SI{4}{\hour} of cool-down~\cite{Tchelidze2019}. Moreover, as mentioned before, several hardware modifications will need to be applied on the ESS LINAC in order to make it able to produce the intense neutrino beam. The modification programme includes, but is not limited to: the upgrade of the low energy beam transport (LEBT), the medium energy beam transport (MEBT), the radiofrequency quadrupole (RFQ), the drift-tube linac (DTL) tank, the Modulator Capacitor and the Cooling System.

\subsection{\textbf{Accumulator Ring}}
\label{accumulator}

The underground accumulator ring (AR), which has a circumference of about \SI{380}{\meter}, will receive the four ca \SI{0.79}{\milli\second} long sub-pulses separated by ca \SI{100}{\nano\second}, which is the time needed to raise the field in the extraction kicker between the sub-pulses. Each sub-pulse will be injected during ca 600 turns and then extracted in one turn, thus producing four ca \SI{1.2}{\micro\second} long proton pulses separated by almost \SI{0.9}{\micro\second} that will each be sent to one of the four separate targets. The H$^{-}$ pulse, which will be injected into the AR via a transfer line from the LINAC, will be stripped at the entrance of the AR using thin carbon foils. The temperature to which these foils are heated must be kept below \SI{2000}{\kelvin} above which the foil sublimation rate will be too high. To achieve this, the incoming beam will be scanned (painted) laterally over the foil, to spread the heating over the foil surface. As a round beam cross-section is desirable when the beam hits the target, anti-correlated painting of the beam has been opted for. Four thinner foils, stacked one behind the other, will be used rather than a single thicker foil. It is planned to investigate, as an alternative to foil stripping, the use of a laser beam to strip the incoming H$^{-}$ ions. This method is currently being developed at the SNS in the US.\\

\begin{figure}[h]
  \begin{center}
    \includegraphics[width=0.59\linewidth]{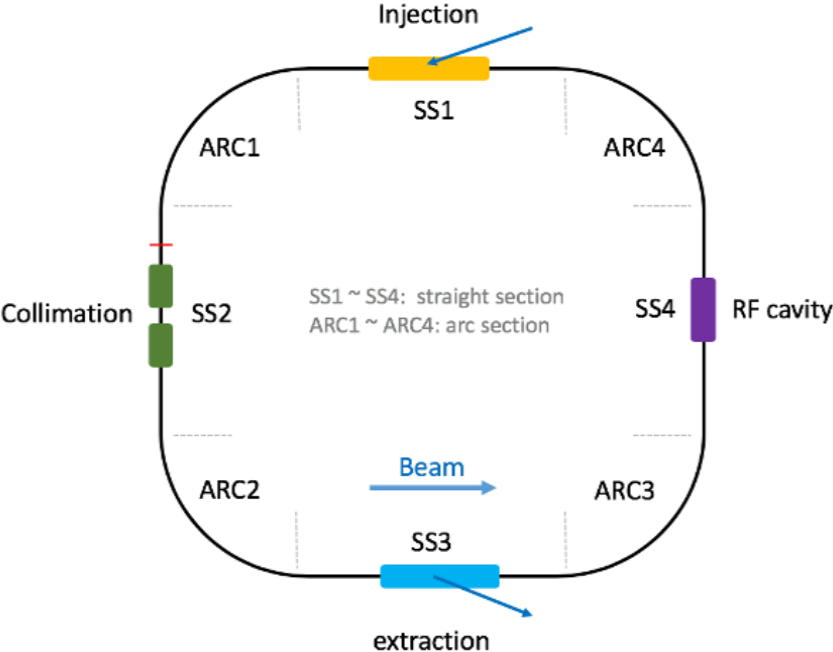}
\caption{The accumulator ring layout.}
    \label{fig:accumulator_layout}
  \end{center}
\end{figure}

The design of the accumulator beam lattice has been carried out using both, beam optics and multi-particle simulations. The simulations show that a geometric 100$\%$ emittance as low as 60 $\pi$ mm mrad is achievable. The total tune spread expected is around 0.05, which means that space charge is not a problem for the AR design. We chose an accumulator layout that has four rather short arcs connected by relatively long straight sections, see Figure~\ref{fig:accumulator_layout}. The arcs contain four Focusing-Defocusing (FODO) cells, each with two dipole magnets, with the dipole magnet centered between two quadrupole magnets. The number and length of the dipole magnets have been chosen to reach the desired bending radius using a moderate magnetic field strength of \SI{1.3}{\tesla}. The main challenges of the design at present are to control the beam loss and to find a H$^{-}$ stripping scheme that is reliable over time. The design of a two-stage collimation system has been made to meet these challenges. A barrier RF cavity will be used to contain the beam pulses longitudinally and preserve the \SI{100}{\nano\second} particle-free gap required for extraction. The beam will be extracted from the ring using a set of vertical kicker magnets and a horizontal septum and the four sub-pulses will be guided by a \SI{72}{\meter} transfer line and a switchyard to the four separate targets.\\

The design work on the Accumulator consists of making the design of a transfer line, an accumulator ring and a switchyard in order to transport and transform the long pulses of H$^-$ ions from the LINAC to short pulses of protons to the target, with minimum losses. Figure~\ref{fig:ESSBuild_Layout} shows the location of the R2S beam-line, AR, the switchyard and the target station.

\subsection{\textbf{Beam Switchyard}}
\label{BSY}
\begin{figure}[h]
\begin{center}
\includegraphics[width=0.66\linewidth]{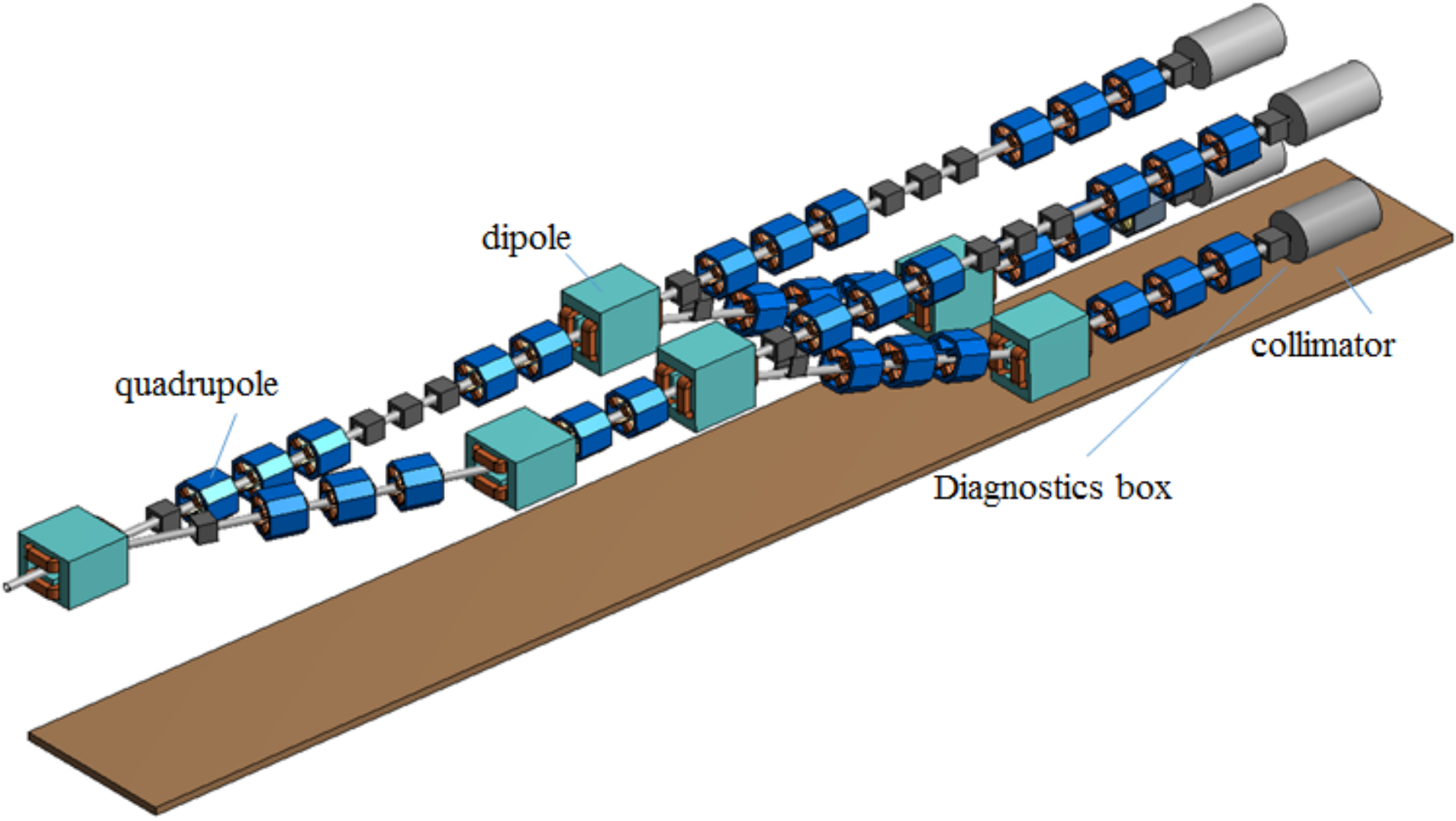}
\caption{\small A 3D isometric view of the beam switchyard.}
\label{fig5.9}
\end{center}
\end{figure}

After their extraction from the accumulator, the protons  will propagate through a beam-line up to a beam switchyard (BSY), in order to be distributed onto the four targets. This beam-line, called R2S, will be mainly composed of quadrupoles and dipoles. One of the main requirements to design the R2S beam-line is to bring the protons into the right direction of the neutrino beam, in the horizontal and vertical planes. According to the simulations, and having taken into account all the requirements, the R2S is foreseen to be \SI{72}{\meter} long. The BSY has to comply with the time structure of the beam extracted from the accumulator. According to the baseline, the accumulator will deliver every \SI{72}{\milli\second} a bunch of four pulses of \SI{1.2}{\micro\second} spaced by almost \SI{0.9}{\milli\second}, that will hit the four targets one after the other. Figure~\ref{fig5.9} shows a 3D CAD drawing of the proposed BSY. The system has a total length of \SI{45}{\meter}. Moreover, several operating failure scenarios of the BSY have been studied. 

\subsection{\textbf{Target Station}}
\label{targetstation}

\begin{figure}[h!]
\begin{center}
\includegraphics[width=1\linewidth]{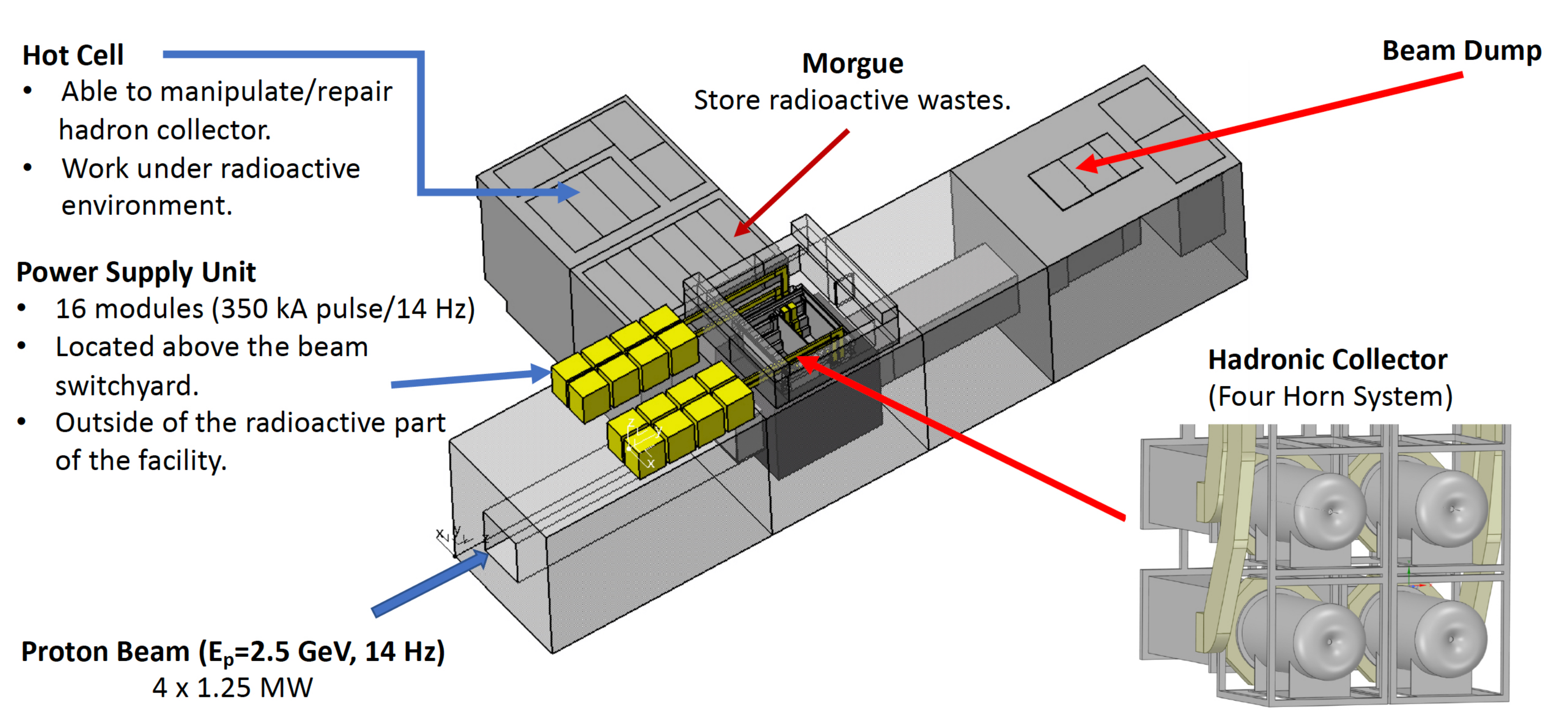}
\caption{Overview of the target station complex with a zoomed view of the 4-horn system.}
\label{fig:TargetStationOverview}
\end{center}
\end{figure}

\begin{figure}[!htbp]
    \centering
       \includegraphics[width=0.49\linewidth]{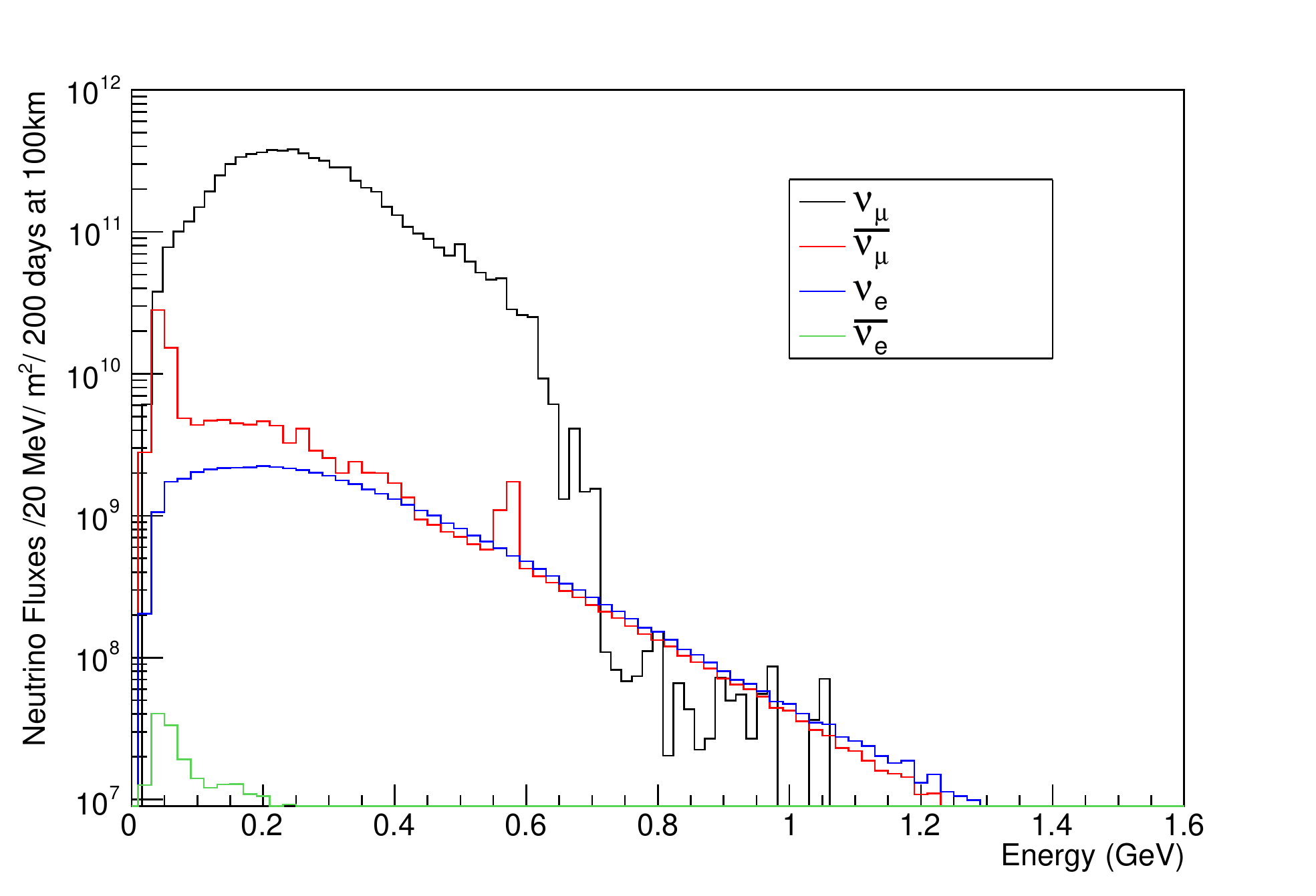}
       \includegraphics[width=0.49\linewidth]{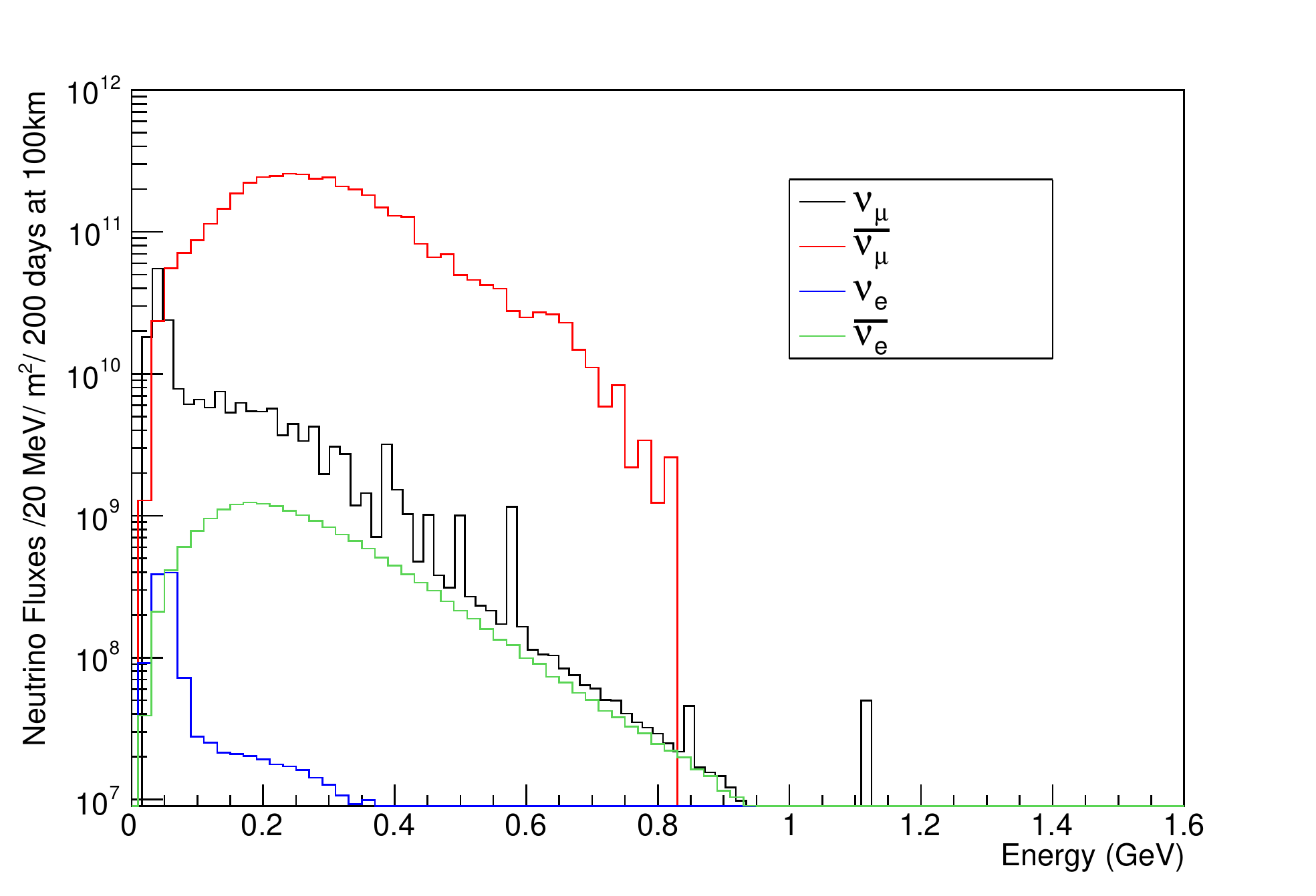} 
    \caption{ESS$\nu$SB neutrino (left) and antineutrino (right) energy spectrum at \SI{100}{\kilo\meter} from the neutrino source.}
    \label{fig:NuFLux1}
\end{figure}

Four identical separate targets will be operated in parallel in order to reduce the beam power that a single target will have to sustain, i.e., from \SI{5}{\mega\watt}/target to \SI{1.25}{\mega\watt}/target. The target design is based on a tube-shaped canister, of ca \SI{3}{\cm} diameter and ca \SI{70}{\cm} length, filled with \SI{3}{\milli\meter} diameter Titanium beads, and cooled using a forced transverse flow of helium gas, pressurised at 10~bar. The primary advantage of such design, in comparison to the monolithic targets, lies in the possibility of making the cooling medium flow directly through the target and by doing so allow a better heat removal from the target areas of highest power deposition. Each target is surrounded by a pulsed magnetic horn (mini-Boone type \cite{osti_900360}) providing a strong toroidal field, with a value of the magnetic field strength of $B_{max} = \SI{1.97}{\tesla}$ at peak current. This is required for the focusing of the charged pion beam, produced from the interaction between the impinging protons on the Ti target in the forward direction into a \SI{50}{\meter} long decay tunnel filled with He gas to reduce interactions. The charge sign of the pions being focused will be changed by reverting the direction of the current in the horn. Each horn will have a separate power supply-unit capable of providing a \SI{350}{\kilo\ampere} current pulse with a flat top of \SI{1.3}{\micro\second} that will be sent to the horn through strip lines. Several horn designs were investigated in this study, of which the Van der Meer horn structure \cite{vanderMeer:278088} showed the best performance. Figure~\ref{fig:TargetStationOverview} shows a 3D CAD drawing of the target station complex with a zoom-in on the 4-horn system. The geometry of the horn has been designed and optimised using a so-called genetic algorithm to provide optimal signal efficiency in the Far Detector. Figure~\ref{fig:NuFLux1} (right) left shows the (anti) neutrino flux distribution at a \SI{100}{\kilo\meter} distance from the neutrino source, resulted from the horn and decay tunnel geometry optimisations.\\

The magnetic field in each horn is produced by a Power Supply Unit (PSU), which delivers the current pulses to the four horns, synchronised with the proton beam pulses coming from the switchyard. The PSU unit consists of 16 modules connected in parallel and able to deliver \SI{350}{\kilo\ampere} with \SI{100}{\micro\second} time width pulse at \SI{14}{\hertz} to each horn. Each proton pulse will deliver $2.23 \times 10^{14}$ protons per target and will have a total energy of \SI{89}{\kilo\joule}. Due to the high power and short pulse duration of the proton beam, the ESS$\nu$SB target will be operating under severe conditions. The energy deposition and the resulting displacement per atom (DPA) in the target have been studied using Fluka. The simulations show a total energy deposition of \SI{0.276}{\giga\electronvolt}/pot/pulse, corresponding to \SI{138}{\kilo\watt}, in the target body. The results of the DPA analysis show that ca 8~DPA/yr are produced along the target. If we consider 1~DPA as a reference value, to estimate the lifetime of the target, the previous value of the DPA corresponds to $ 3.024 \times 10^{7} $ pulses.\\

At the end of the decay tunnel there will be a water-cooled beam-dump (BD) that will absorb the non-interacted primary proton beam and the undecayed muons. The BD core structure will consists of four independent graphite blocks, each block facing one of the four horns. The four segments are supported on a cross-like structure, made of a Copper-Chromium-Zirconium (CuCr1Zr-UNS C18150 \cite{CuAlloy}) alloy, similar to that used for the new PSB \cite{PSBBeamDump} and ESS beam dumps. Each segment is in turn constructed from two zigzag blocks with \SI{1}{\cm} opening between them, to allow for thermal expansion of the individual blocks. The \SI{30}{\cm} support plates are used as heat sinks with water channels drilled inside the plate body.\\

\begin{figure}[h!]
\begin{center}
    \centering 
    \includegraphics[width=0.43\linewidth]{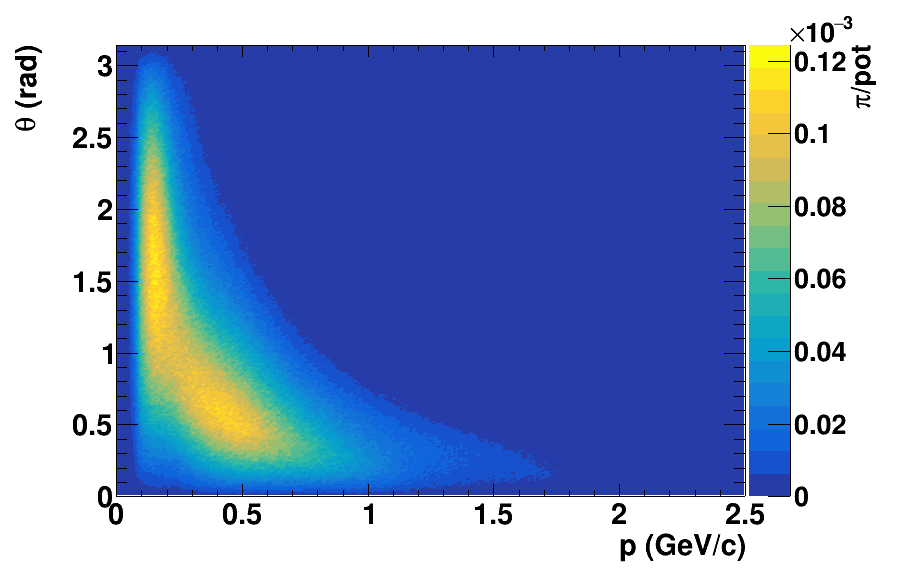}
    \includegraphics[width=0.43\linewidth]{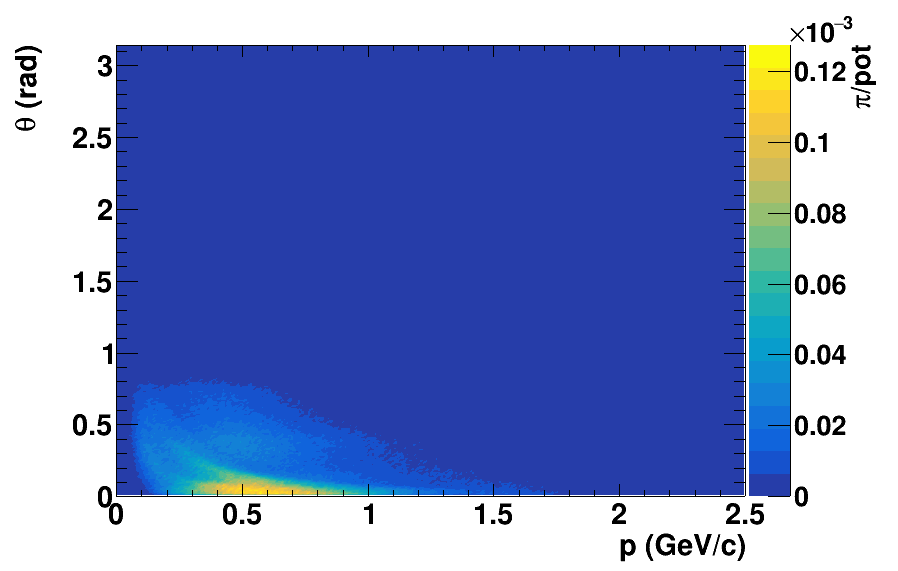}
\caption{Angular-momentum distribution of the positive pions, exiting the target (left) and entering the decay tunnel after focusing (right).}
\label{fig:Esb_Pions}
\end{center}
\end{figure}

Due to the comparatively low energy of the ESS LINAC protons, the pions exiting the target will have a wide angular and momentum distribution, as can be seen from the left panel of Figure~\ref{fig:Esb_Pions}, where angle $\theta$ is the azimuthal angle w.r.t. the proton beam direction. In the right panel of the same figure, the distribution of the pions entering the decay tunnel, after focusing, is shown. The yield of positive pions leaving the target is 0.32 pot$^{-1}$, while that of the negative pions is 0.22 pot$^{-1}$. Of the right-sign pions leaving the target, 50$\%$ enter in the decay tunnel.

\subsection{\textbf{Detectors}}
\label{detectors}

The detector complex of ESS$\nu$SB comprises a Near Detector (ND) and a Far Detector (FD). 

\subsubsection {\textbf{Near Detector}}

\begin{figure}[!htbp]
    \centering
       \includegraphics[width=0.48\linewidth]{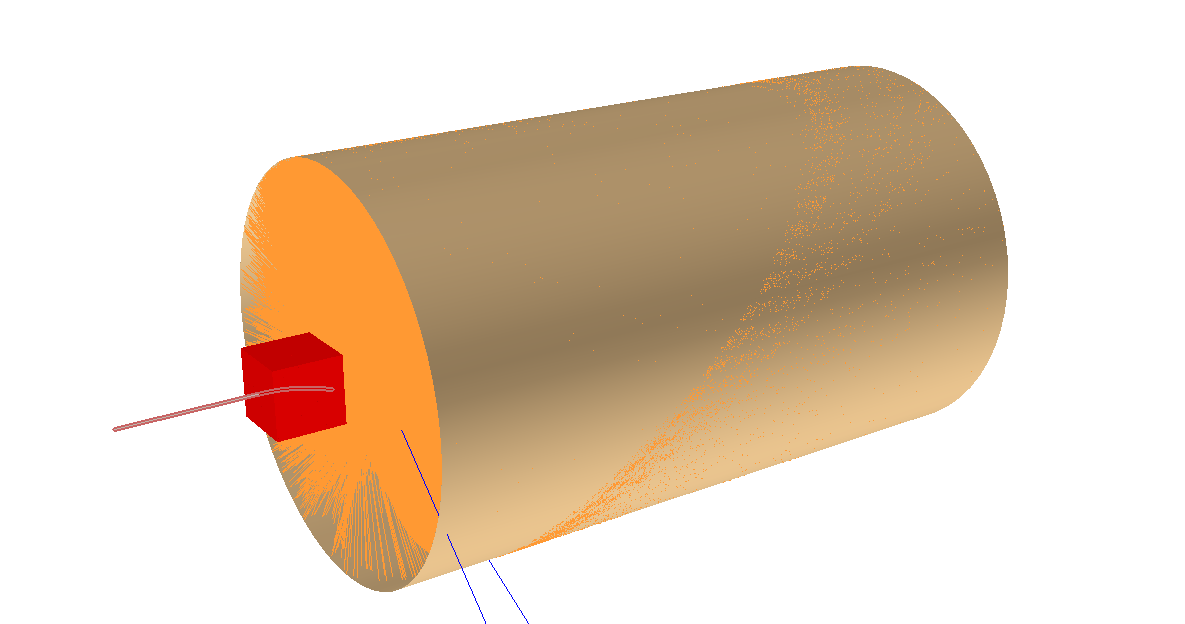}
       \includegraphics[width=0.5\linewidth]{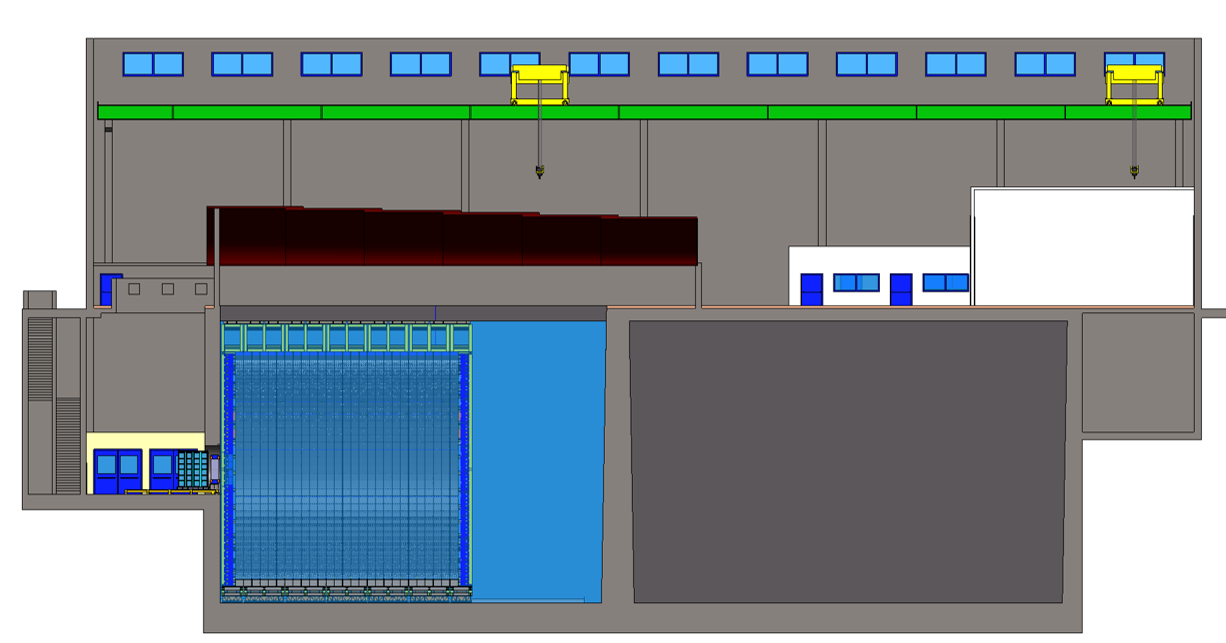} 
    \caption{The ESS$\nu$SB near detector layout. In the left picture, an artistic view of the detector is shown with an event happened in the sFGD. The trajectory of a positive muon (red) in the SFGD (bended by the magnetic field) and the two neutrinos (blue) from its decay in the WatCh are shown. Emitted Cherenkov photons in the WatCh are shown in orange. The right picture, presents an engineering design of the detector and the cavern.}
    \label{fig:detectors:NDcomplex}
\end{figure}

The purpose of the ND is to monitor the neutrino beam intensity and to measure the muon- and electron-neutrino cross-sections, in particular their ratio, which is important for minimizing the systematic uncertainties in the experiment. The ND will be located underground within the ESS site ca \SI{250}{\meter} from the target station. It will be composed of three coupled detectors: A kiloton mass \textbf{Water Cherenkov detector (WatCh)}, which will be used for event rate measurement, flux normalisation and for event reconstruction comparison with the FD, a \textbf{magnetised super Fine-Grained Detector (sFGD)} \cite{Sgalaberna:2017}, located inside a \SI{1}{\tesla} dipole for measurements of the poorly known neutrino cross sections in the energy region below \SI{600}{\mega\electronvolt}, and placed upstream of, and adjacent to the water volume, and \textbf{an emulsion detector} of similar type as in the NINJA experiment \cite{Hiramoto:2013}. Figure~\ref{fig:detectors:NDcomplex} (right) shows the layout of the ND complex.

\begin{figure}[!htbp]
    \centering
     \includegraphics[width=0.6\linewidth]{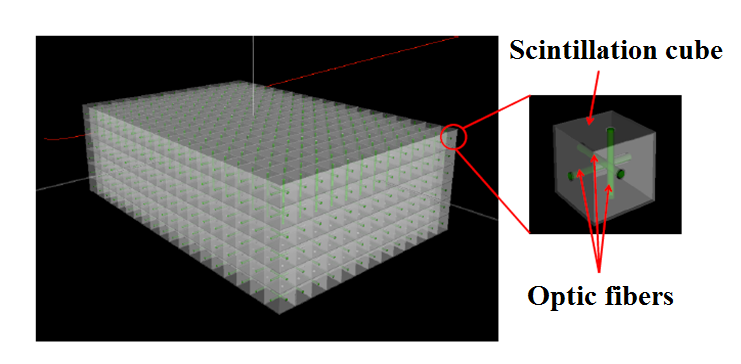}
     \includegraphics[width=0.35\linewidth]{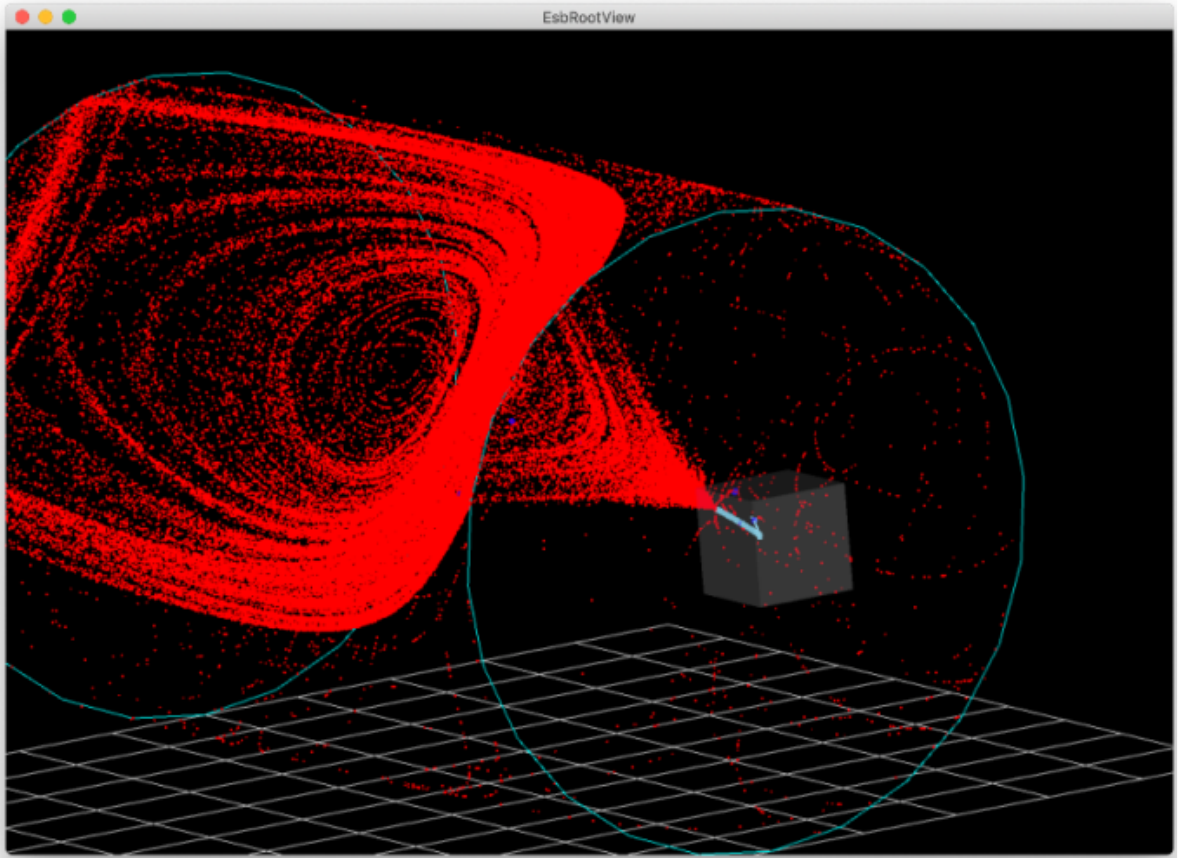}
    \caption{(Left) design of the Super Fine-Grained Detector with three-dimensional read-out. (Right) An $\nu_{\mu}$ interaction event in the SFGD cube with a secondary muon producing a Cherenkov light in the near water Cherenkov detector.}
    \label{fig:detectors:SFGD-1}
\end{figure}

The WatCh water-tank will be a horizontal cylinder of $\sim$~\SI{11}{\meter} length and \SI{4.72}{\meter} radius (total volume $767 m^3$) with the inner walls having a 30\% coverage of 3.5~inch Hamamatsu R14689 photomultiplier tubes (PMTs). The WatCh detector consists of modules, each of them housing 16 PMTs. There are in total 1258 modules altogether containing 20128 PMTs. The 1.4 $\times$ 1.4 $\times$ 0.5~$m^{3}$ sFGD, located in front of the water Cherenkov detector, will be composed of $10^{6}$ \SI{1}{\cm}$^{3}$ plastic scintillator cubes read out by three-dimensional pattern of wave-length-shifting optical fibres (Figure~\ref{fig:detectors:SFGD-1} (left)). The thickness of \SI{0.5}{\meter} along the beam axis is chosen in order to have a sufficient number of charged leptons that continue into the WatCh and thus to allow the combination of the information from the two  detectors. A magnetic field of up to \SI{1}{\tesla} and perpendicular to the beam is applied in the tracker by a dipole magnet. In front of the tracker, a NINJA type emulsion detector is situated that will be used for cross-section measurements. The NINJA detector is an emulsion-based detector with a water target, currently under operation as one of the T2K near detectors. Its primary purpose is to  measure precisely the neutrino interaction topology and double differential cross-sections. Members of the NINJA Collaboration are prepared to add a similar detector to the ESS$\nu$SB suite of the near detector.\\

\noindent\textbf{sFGD and WatCh combined analysis}

Figure~\ref{fig:detectors:SFGD-1} (right) shows an example of what is here called a cross-over event in which a muon neutrino interaction in the sFGD and the secondary muon continues into the WatCh and produces Cherenkov light. Around 12\% (20\%) of positive (negative) muons produced in the sFGD will continue into the WatCh and be detected there. For electron neutrino interactions such events represent about 6\% of the sample. For the crossover events we aim at a good purity of the electron neutrino event sample by efficiently rejecting events originated in the sFGD that have muons continue into the WatCh by exploiting the sFGD and WatCh ID capabilities together, which will be used for electron neutrino cross-section measurement.

 \begin{figure}[!htbp]
    \centering
     \includegraphics[width=0.48\linewidth]{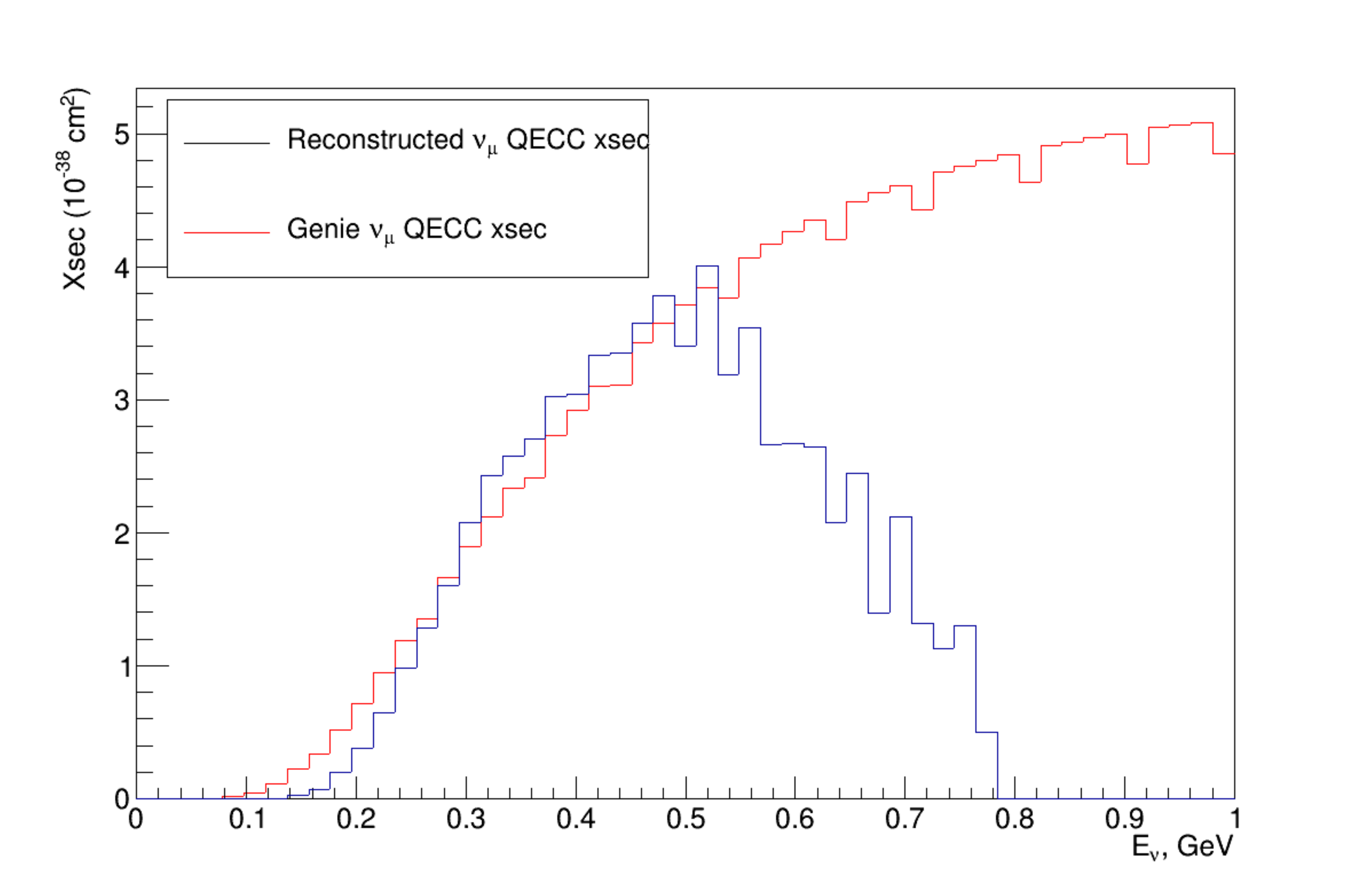} 
     \includegraphics[width=0.48\linewidth]{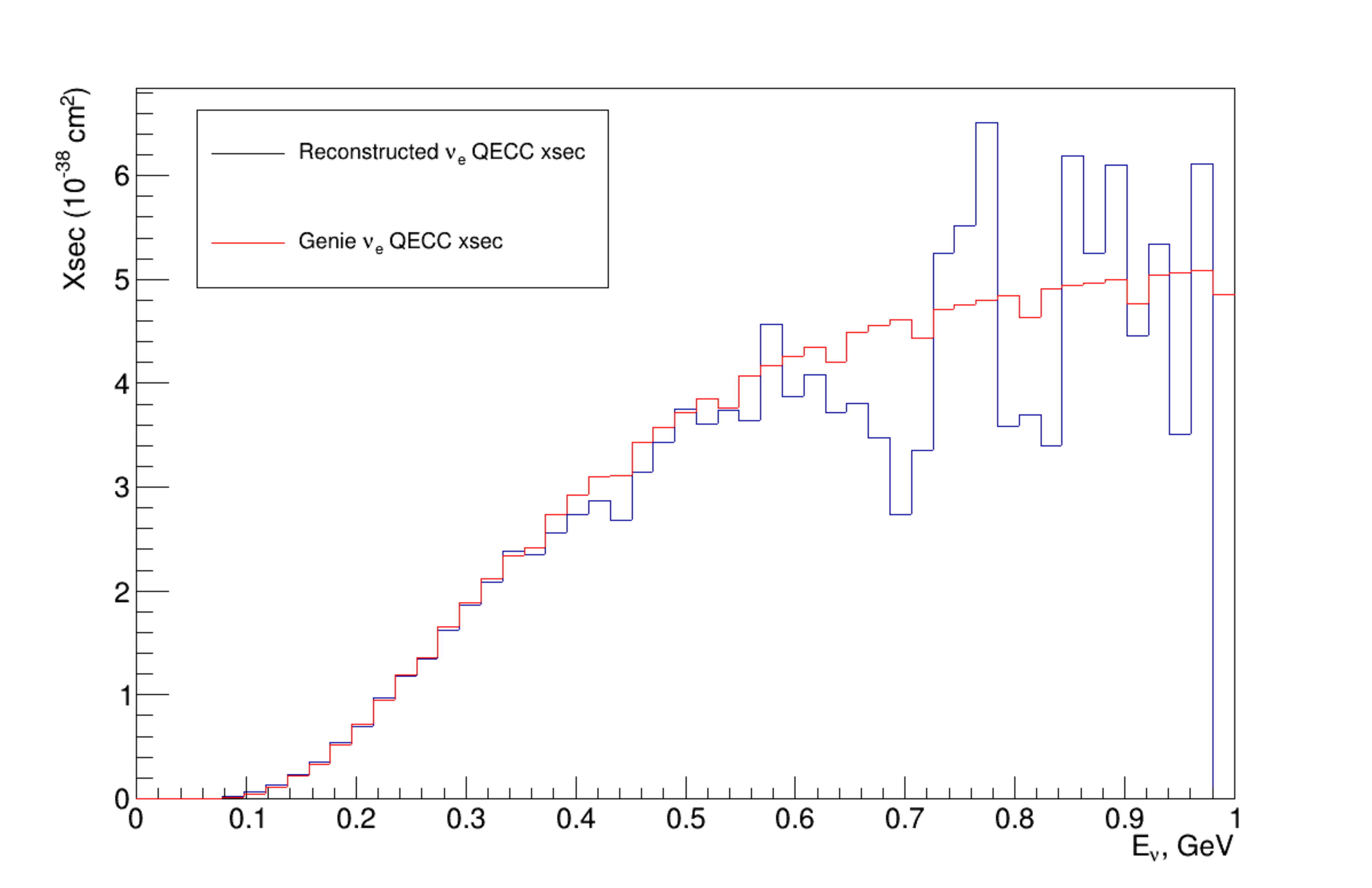} 
     \caption{Left: "Measured" $\nu_\mu$ cross-section of (blue) compared to the cross-section used by GENIE (red) to simulate the interactions. Muon neutrino energy calculation is based on the fitted muon momentum. Right: "Measured" $\nu_e$ cross-section  (blue) compared to the cross-section used by GENIE (red) to simulate the  interactions. Electron neutrino energy calculation is based on the true electron momentum. }
    \label{fig:detectors:SFGD_xsec}
\end{figure}

\noindent\textbf{Neutrino cross-section measurements at ND}

\begin{figure}[!htbp]
    \centering
     \includegraphics[width=0.65\linewidth]{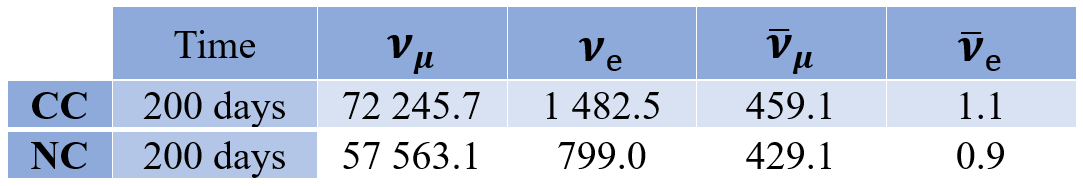}
     \includegraphics[width=0.65\linewidth]{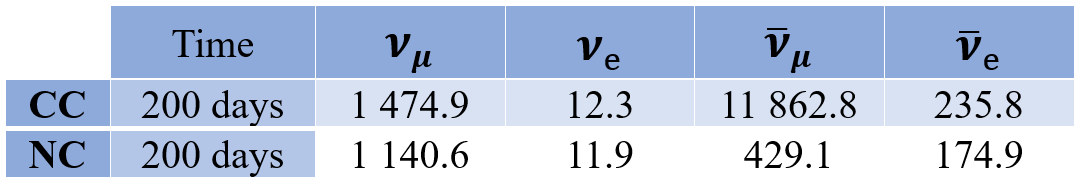}
    \caption{Number of expected interactions in the sFGD: (top) positive horn polarity, i.e. neutrino beam. (Bottom) negative horn polarity, i.e. antineutrino beam.}
    \label{fig:detectors:SFGD_rate}
\end{figure}

As pointed out above, one of the tasks of the sFGD is to measure neutrino cross-sections. In Figure~\ref{fig:detectors:SFGD_xsec}, "measured" $\nu_\mu$ and $\nu_e$ cross-sections are compared with the ones used for simulation of neutrino events. Good agreement up to $E_{\nu} \sim 500$ MeV is observed in both cases. Above that, the discrepancy is big, especially in the muon case. The reason is related to the limited statistics due to the limited number of neutrinos above this momentum in our low-energy neutrino beam. The number of expected (anti)neutrino interactions in the sFGD, obtained by using GENIE neutrino event generator \cite{GENIE:2010, GENIE:2015, GENIE:2021}, is given in the tables in  Figure~\ref{fig:detectors:SFGD_rate} (bottom) top.

\subsubsection {\textbf{The Far Detector}}

The Far Detector, which will detect the rate and energy distributions of the muon- and electron-neutrinos, respectively, will be composed of two vertical cylinders of ca \SI{74}{\meter} in height and ca \SI{74}{\meter} in diameter, installed in caverns ca \SI{1000}{\meter} under the ground level to protect them from the cosmic radiation background. Two locations for the FD are under consideration, both near the position of the second $\nu_\mu-\nu_e$ oscillation maximum, thereby resulting in a majority of the events being collected at the second oscillation maximum. One location is in the Zinkgruvan mine, \SI{360}{\kilo\meter} from the ESS, and the other is in the Garpenberg mine, \SI{540}{\kilo\meter} from the ESS. Both mines are active, presenting the advantage of local services like access shafts and declines, ventilation, drainage and other services that are in operation. The locations of the detector caverns are planned to be outside the region of exploitable ore, but still accessible from the mine drifts and at a sufficient distance (a few hundred meters) from the mining activity areas to avoid any kind of mutual disturbance. The detailed design of the cavern locations must be preceded by core drilling to enable measurements of the rock strength and pressure in the planned underground regions. The choice between the two mines will eventually be made on the basis of such measurements and on the difference in performance for CP violation discovery and CP phase angle measurement accuracy as will be discussed in section~\ref{physicspotential}. The photomultiplier tubes coverage of the ca \SI{25000}{\meter}$^{2}$ inner detector walls will be 40\% (requiring, e.g., ca 50,000 PMTs of 20 inch diameter). The water in the detector tanks will be purified using an industrial-sized water-cleaning plant in order to achieve about \SI{100}{\meter} absorption length for the light wavelength that the PMTs are sensitive to. The ability to dissolve gadolinium salt for increased neutron detection efficiency, and thereby of electron-antineutrinos, will be included.

\section{Physics Potential}
\label{physicspotential}

\begin{figure}[H]
\centering
\includegraphics[width=6.cm]{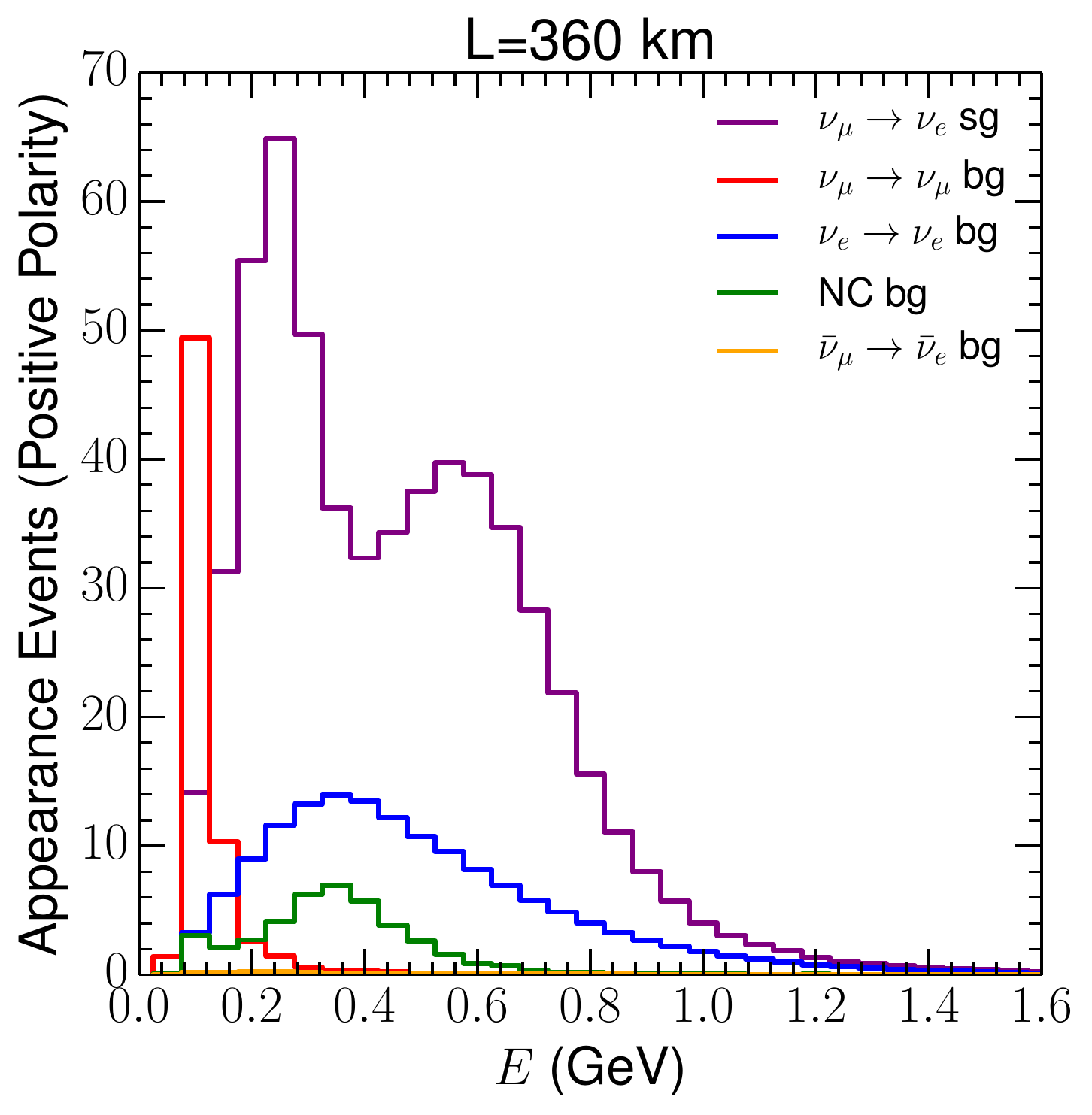}
\includegraphics[width=6.cm]{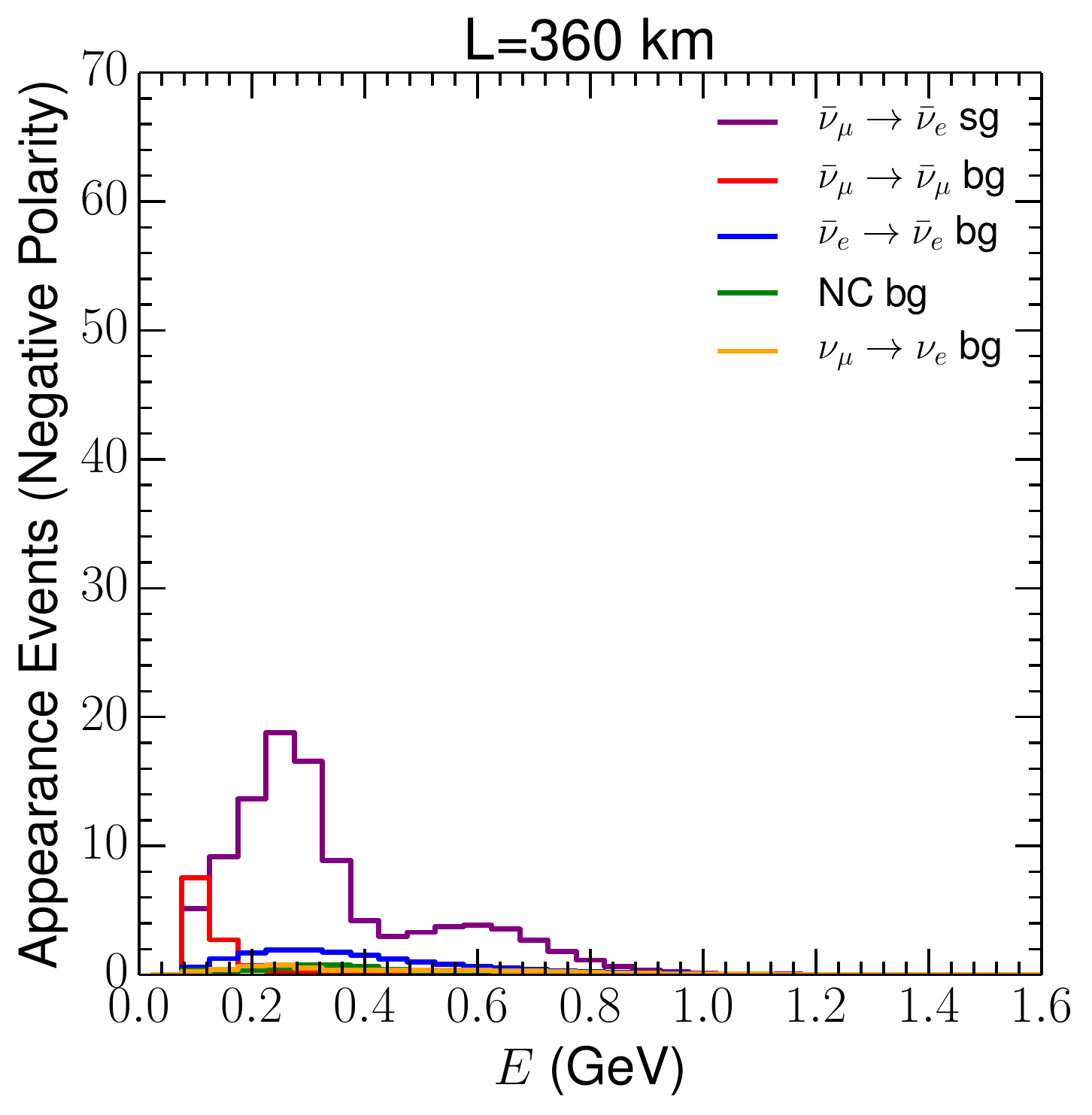}
\includegraphics[width=6.cm]{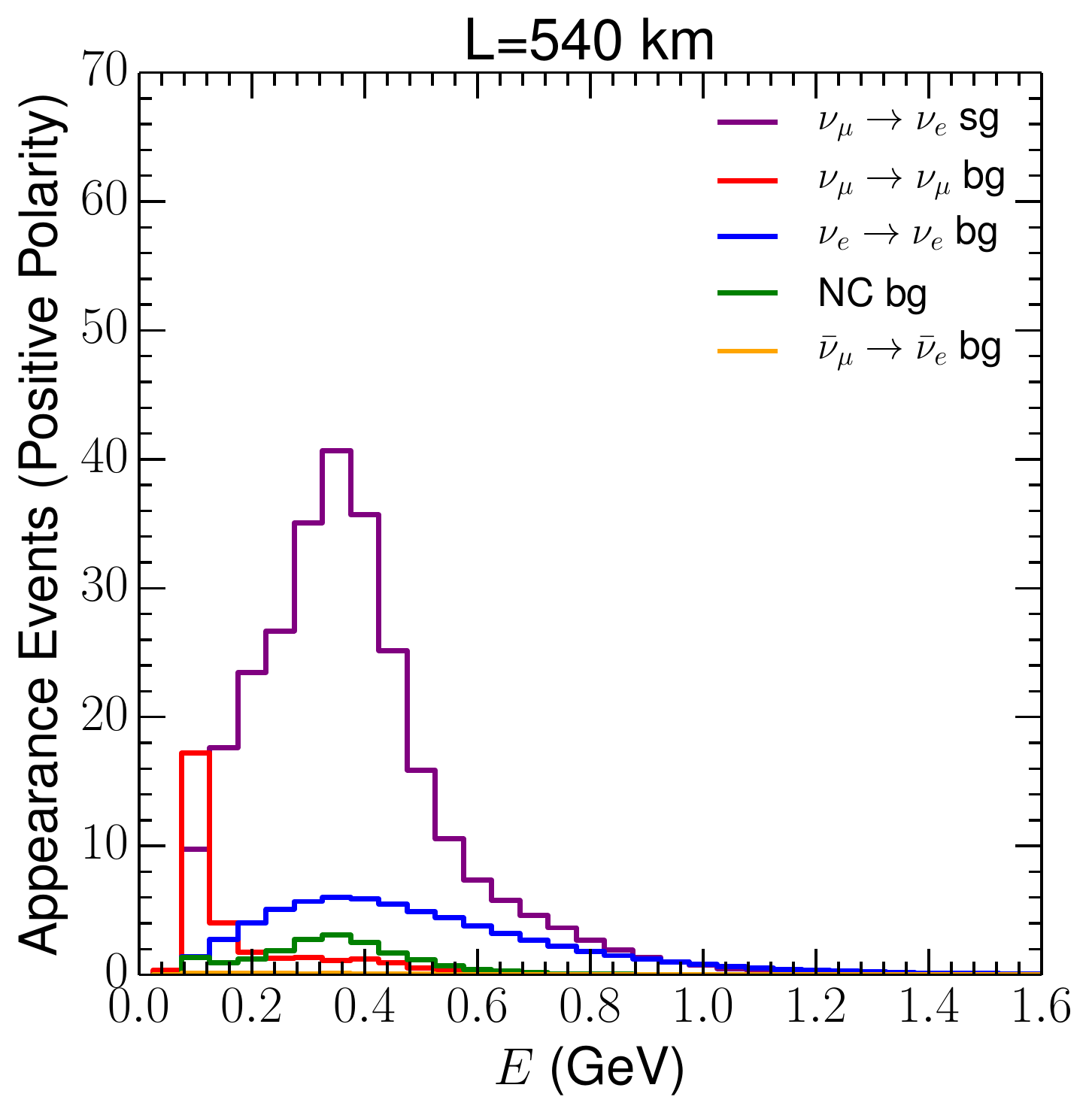}
\includegraphics[width=6.cm]{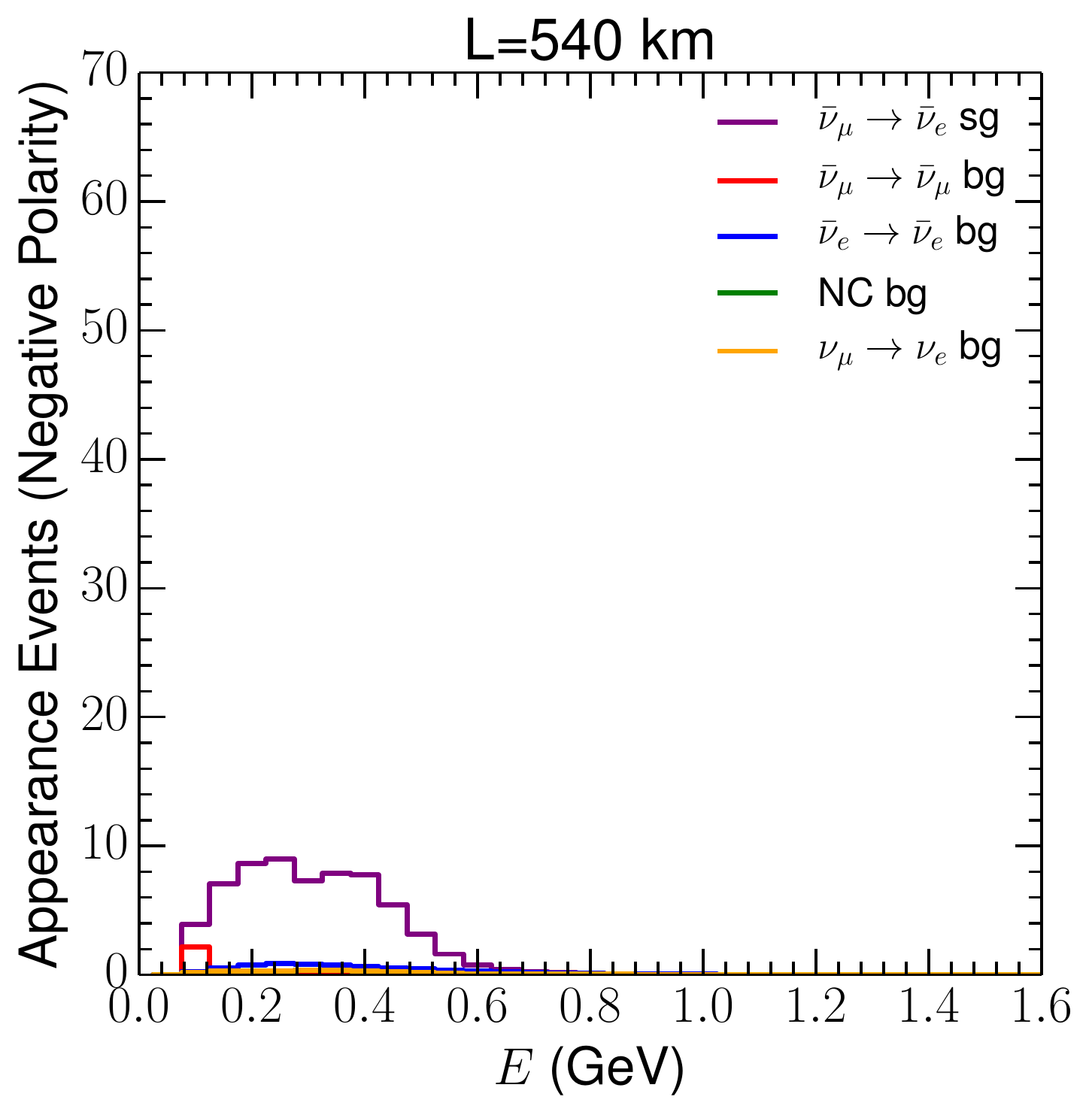}
\caption{Event rates for the different signal and background components for the e-like sample with positive (left panels) and negative focusing (right panels) and for the Zinkgruvan \SI{360}{\kilo\meter} (upper panels) and Garpenberg \SI{540}{\kilo\meter} baselines.}
\label{fig:rates}
\end{figure}

The preliminary optimisation studies presented in Ref.~\cite{Baussan:2013zcy}, as well as the follow-up studies~\cite{Agarwalla:2014tpa,Chakraborty:2017ccm,Chakraborty:2019jlv,Ghosh:2019sfi,Blennow:2019bvl,ESSnuSB:2021azq}, of the physics reach of the ESS$\nu$SB facility for leptonic CP-violation, demonstrate that the best baseline at which to study the neutrino beam would be between \SI{350}{\kilo\meter} and \SI{550}{\kilo\meter}. This makes the ESS$\nu$SB design unique, as the neutrino flux observed by the detector mainly corresponds to the second maximum of the $\nu_\mu \to \nu_e$ oscillation probability, with a more marginal contribution of events at the first oscillation peak. However, as discussed before, there is a price to pay in order to observe the oscillation probability at its second maximum. Even though it is the optimal choice to maximise the dependence of the oscillation probability on the $\delta_{CP}$, the ratio of the oscillation baseline to the neutrino energy ($L/E$) needs to be a factor 3 larger compared to the first maximum. This means that the statistics will be about an order of magnitude smaller than if the detector had been located at the first oscillation peak. Furthermore, the neutrino cross section decreases and beam spread increases for smaller neutrino energies. The \SI{360}{\kilo\meter} baseline option, corresponding to a point between the first and the second ocillation maxima as seen in Figure.~\ref{fig:probs}, represents a compromise between the two choices. The neutrino flux would be 2.25 times larger than that at the \SI{540}{\kilo\meter} option and roughly the same number of events belonging to the second oscillation peak would be observed at either site. At the higher energy end of the spectrum, events from the first oscillation maximum would also be observed for the shorter baseline. This can be seen in the event rates expected at each of the two detector locations depicted in Figure~\ref{fig:rates}. In any case, for the ESS$\nu$SB the choice close to the second oscillation maximum was shown to be optimal for its increased dependence on $\delta_{CP}$ and despite the reduced event rate.

\subsection{\textbf{Physics reach of the ESS$\nu$SB experiment}}

\begin{figure}[H]
\centering
\includegraphics[width=7.5cm]{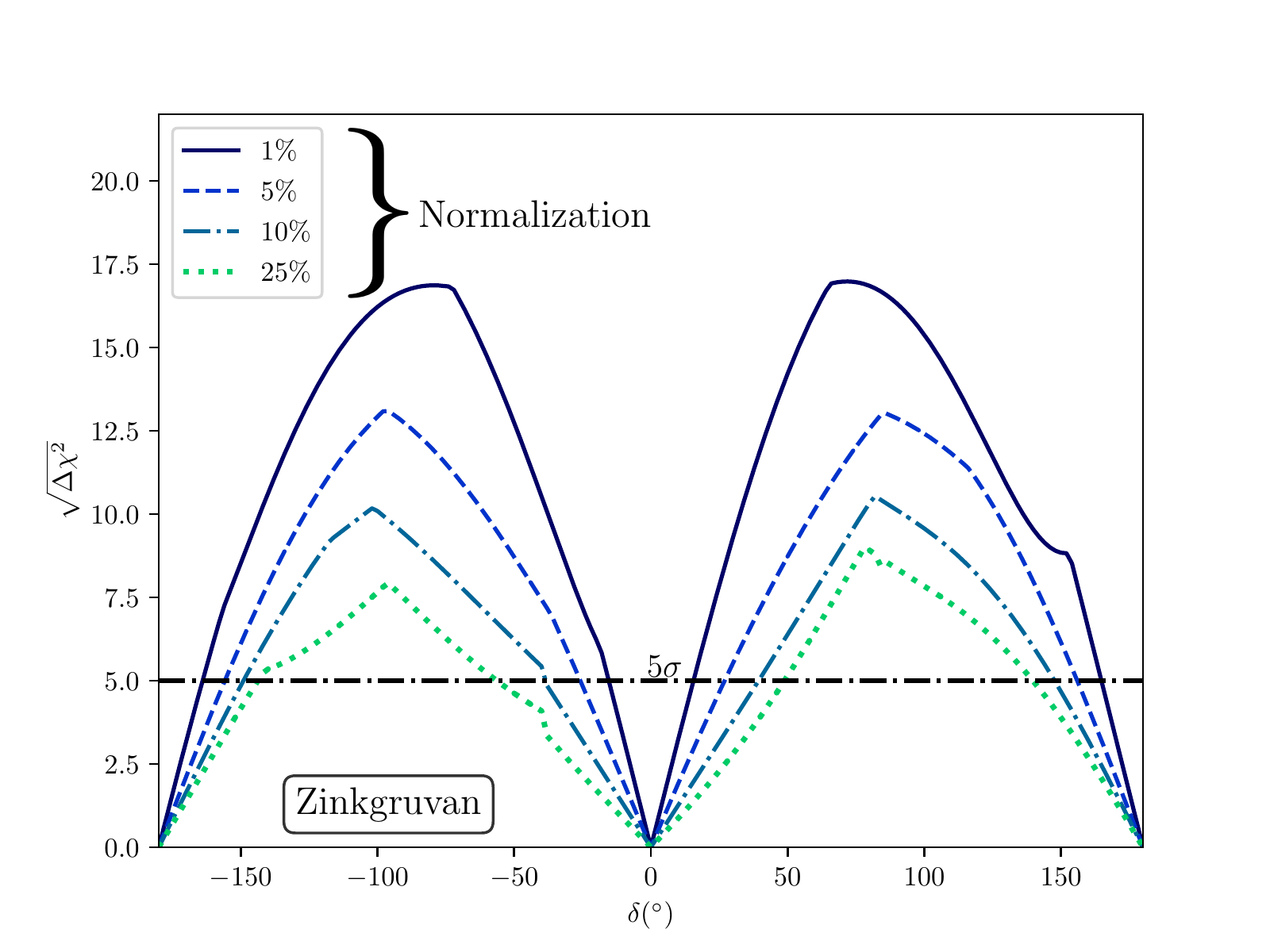}
\includegraphics[width=7.5cm]{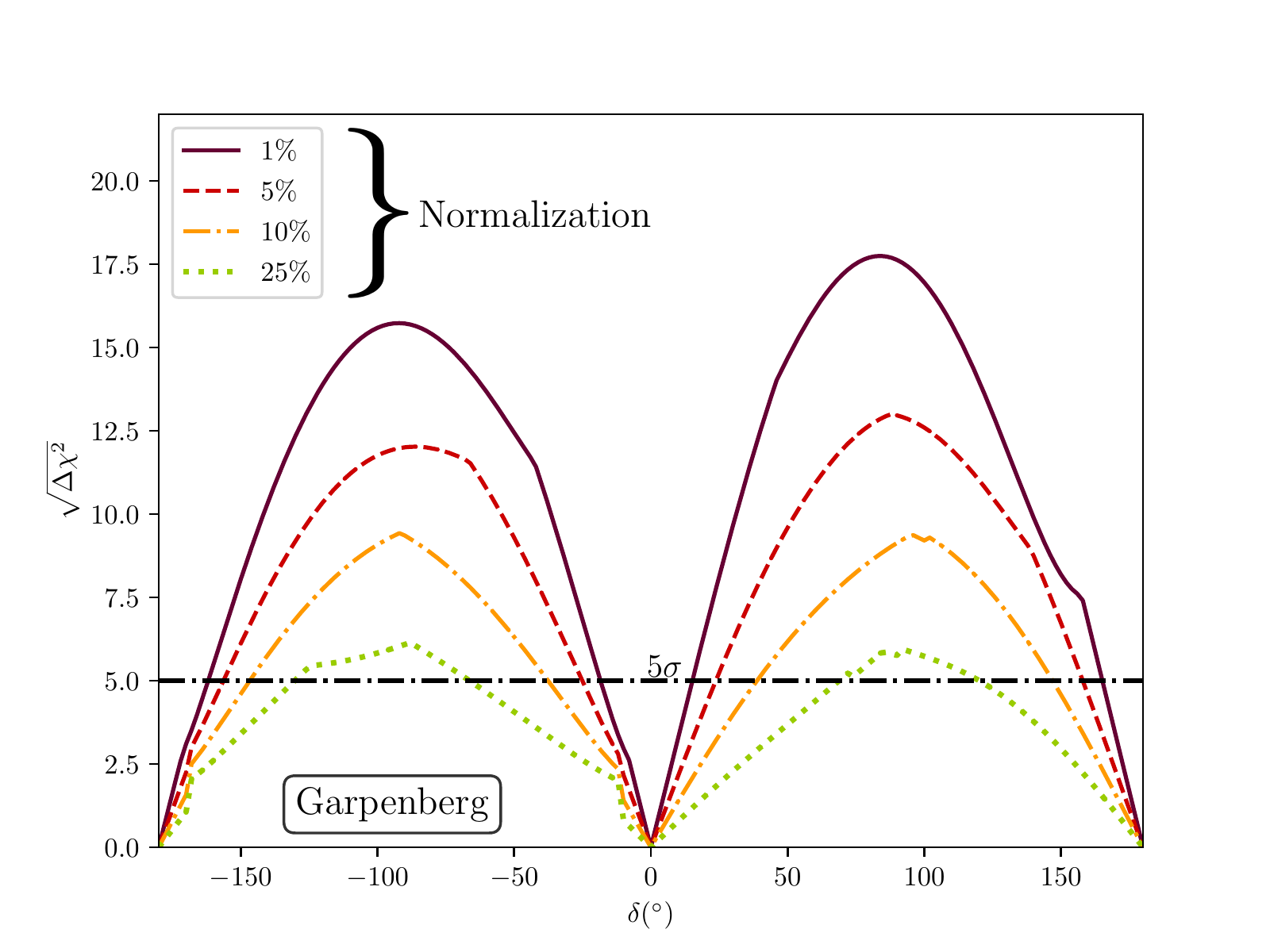}
\includegraphics[width=7.5cm]{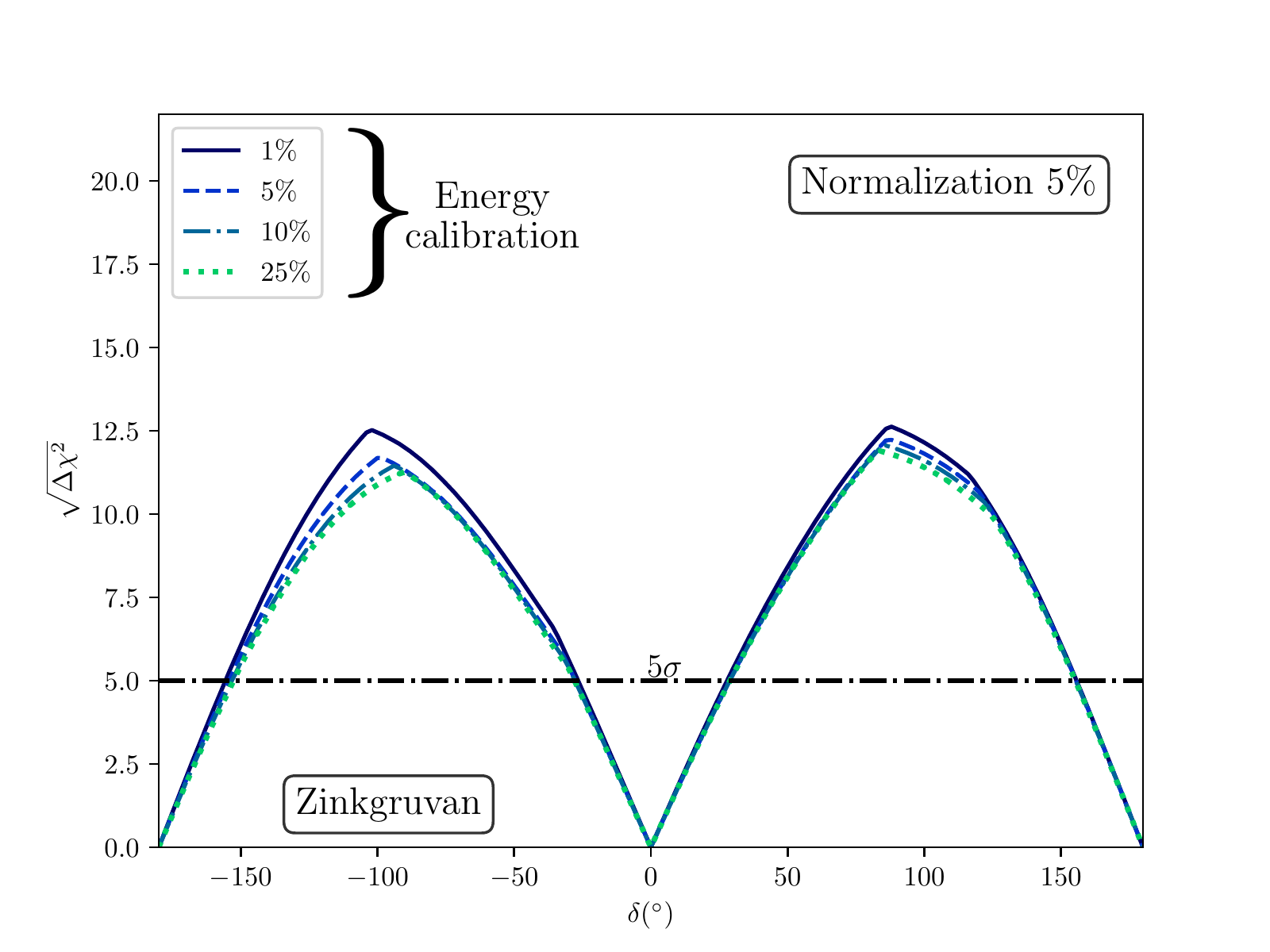}
\includegraphics[width=7.5cm]{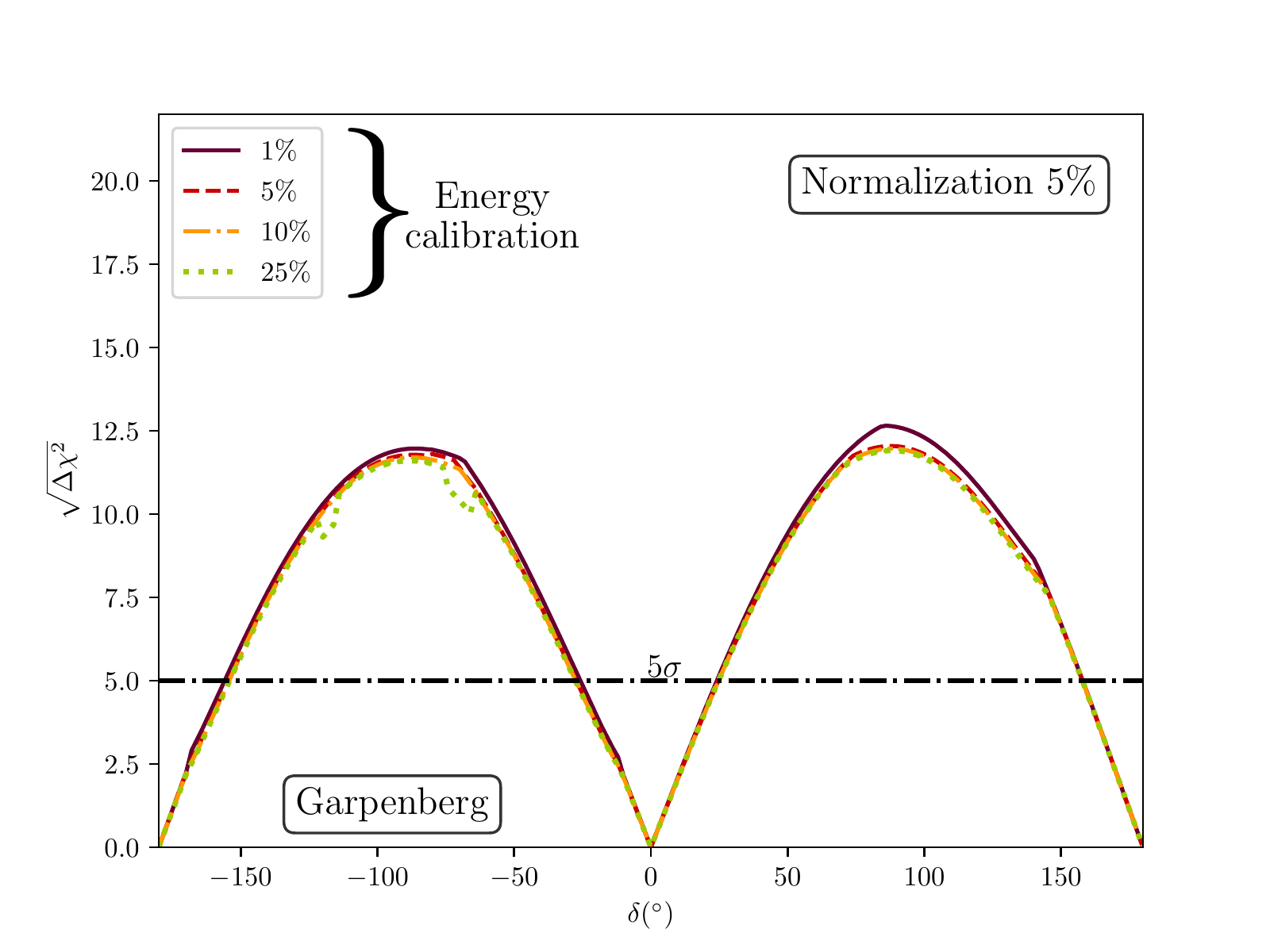}
\includegraphics[width=7.5cm]{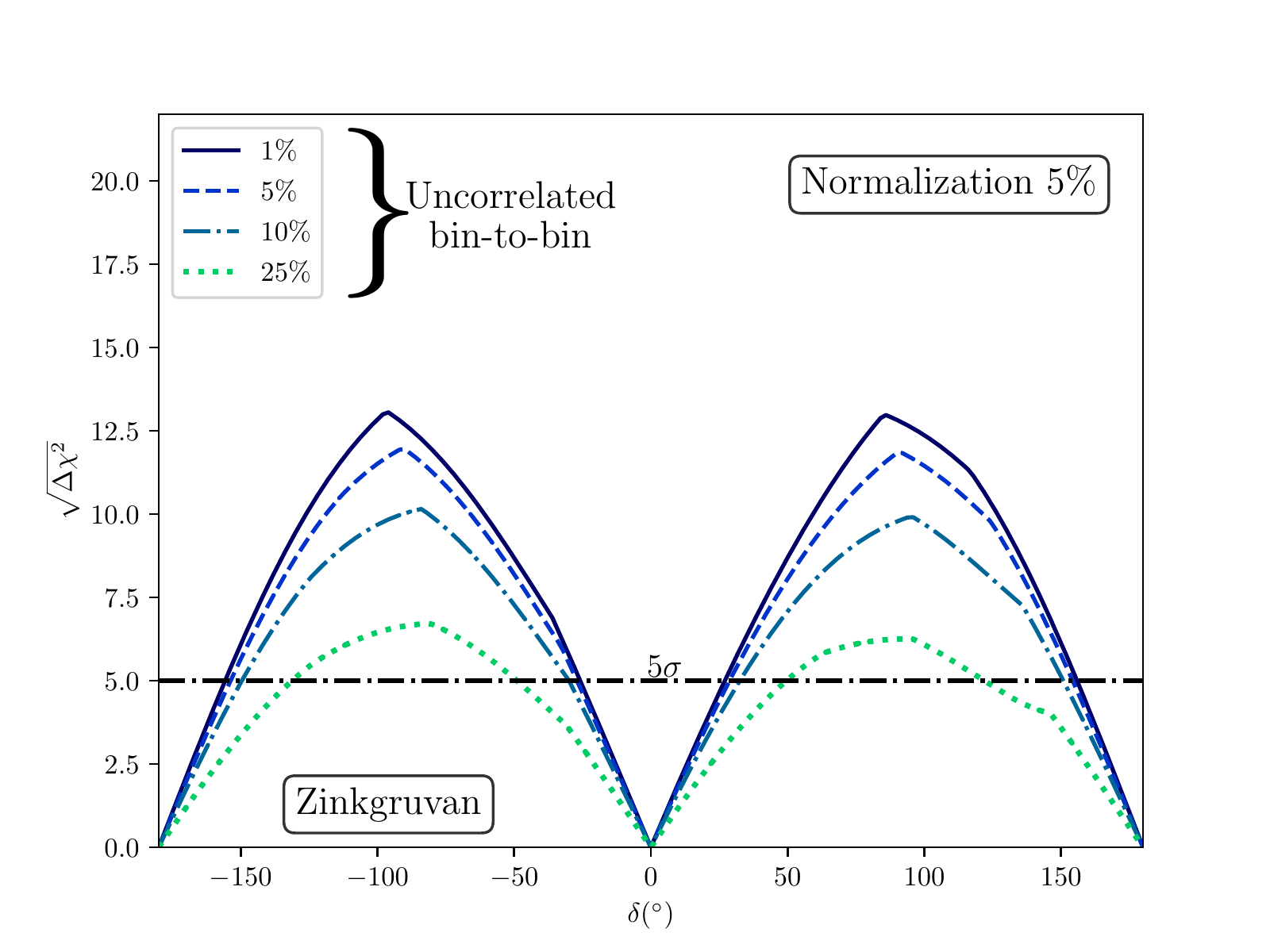}
\includegraphics[width=7.5cm]{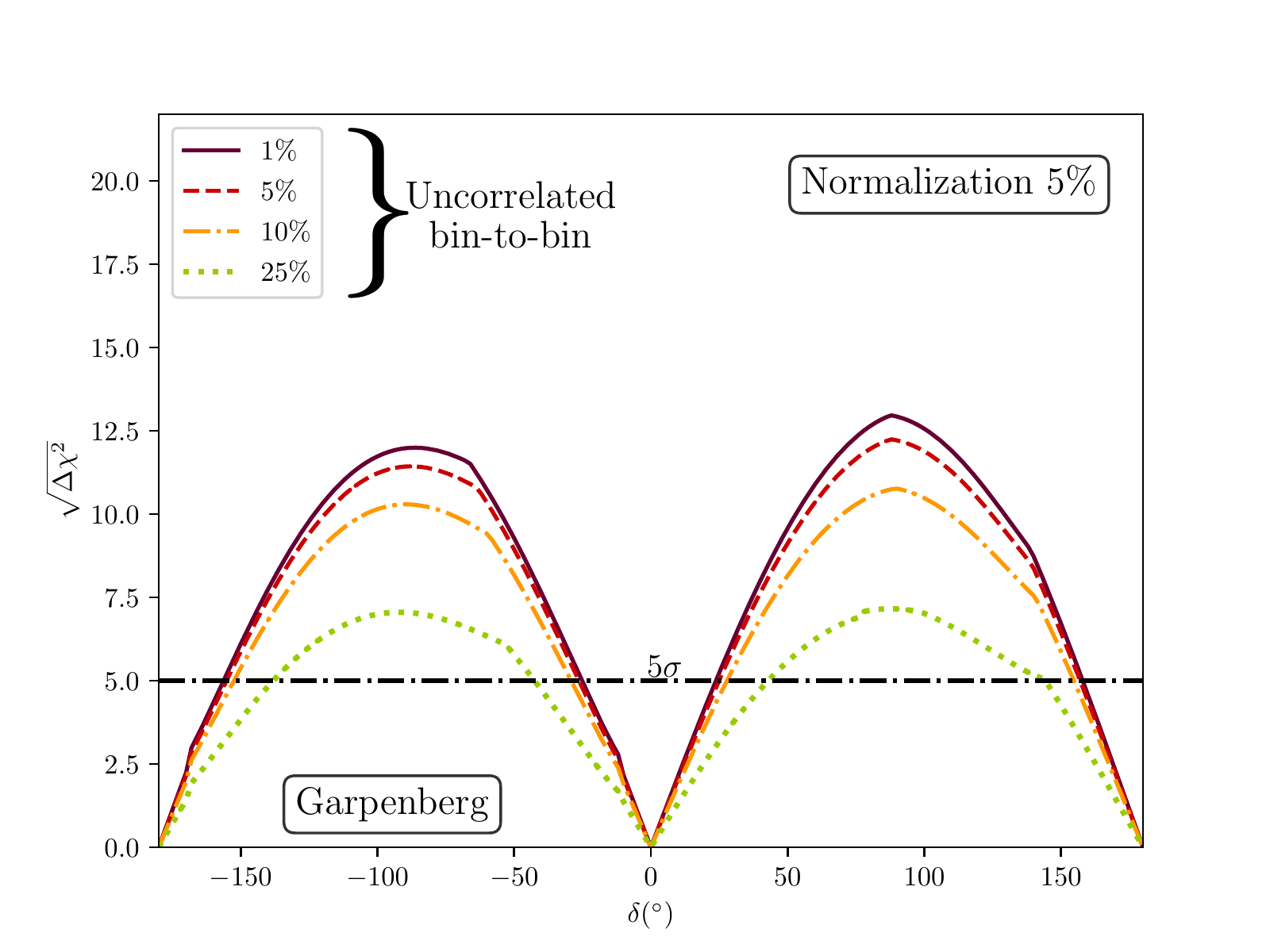} 
\caption{CP discovery potential of the ESS$\nu$SB. Left (right) panels for the Zinkgruvan \SI{360}{\kilo\meter} (Garpenberg \SI{540}{\kilo\meter}) baseline.}
\label{fig:cpsens}
\end{figure}

Figure~\ref{fig:cpsens} shows the CP discovery potential obtained for the two baselines for different assumptions about the size of the systematic uncertainties. Generally, we find that the performance of both baselines is very similar, with only slightly different areas covered above the 5~$\sigma$ mark, depending on the systematic uncertainties considered. In the upper panels, the impact of an overall normalization uncertainty, uncorrelated among the different signal and background samples of $1\%$, $5\%$, $10\%$ and $25\%$ uncertainties, is shown. It is remarkable that, even for an extremely large uncertainty value of $25\%$, a significant portion of the values of $\delta_{CP}$ would still allow a discovery of CP violation above the 5~$\sigma$ level. In the simulation, this uncertainty is uncorrelated between the neutrino and antineutrino samples and therefore is able to mimic the effect of CP violation. Nevertheless, a discovery would be possible even in this scenario. This can be understood from Figure~\ref{fig:probs} where, close to the second oscillation peak, changes in $\delta_{CP}$ can lead to changes in the probability, even above the $25\%$. The dependence of the shape of the oscillation probability on the value of $\delta_{CP}$ may also contribute to the sensitivity. In the middle and lower panels of Figure~\ref{fig:cpsens}, we also explore how robust the results obtained are against other systematic uncertainties that might affect the shape of the measured spectrum. In particular, in the middle panels, the impact of a $1\%$, $5\%$, $10\%$ and $25\%$ uncertainty in the energy calibration is shown. In the lower panels, a more general bin-to-bin uncorrelated set of nuisance parameters has been considered. In both cases, a $5\%$ normalization uncertainty has been added to allow the possible interplay between the different sets of systematic uncertainties that may be relevant. We find that the energy calibration uncertainty has a rather minor impact in the CP discovery potential for both baselines under study. Conversely, the more general implementation of uncorrelated uncertainties in each bin can have a more significant impact, but still a smaller one than the overall normalization considered in the upper panels. The results demonstrate that the ESS$\nu$SB setup has remarkable CP discovery potential even for very conservative assumptions on the systematic uncertainties that could affect the far detector. Indeed, the measurements of the near detector will keep these uncertainties at the few $\%$ level, for the present generation of neutrino oscillation experiments. In particular, assuming a $5\%$ normalization uncertainty in line with assumptions made for similar facilities, we find that CP violation could be established for a $71\%$ ($73\%$) of the values of $\delta_{CP}$ for the \SI{360}{\kilo\meter} (\SI{540}{\kilo\meter}) baseline.

\begin{figure}[H]
\centering
\includegraphics[width=7.5cm]{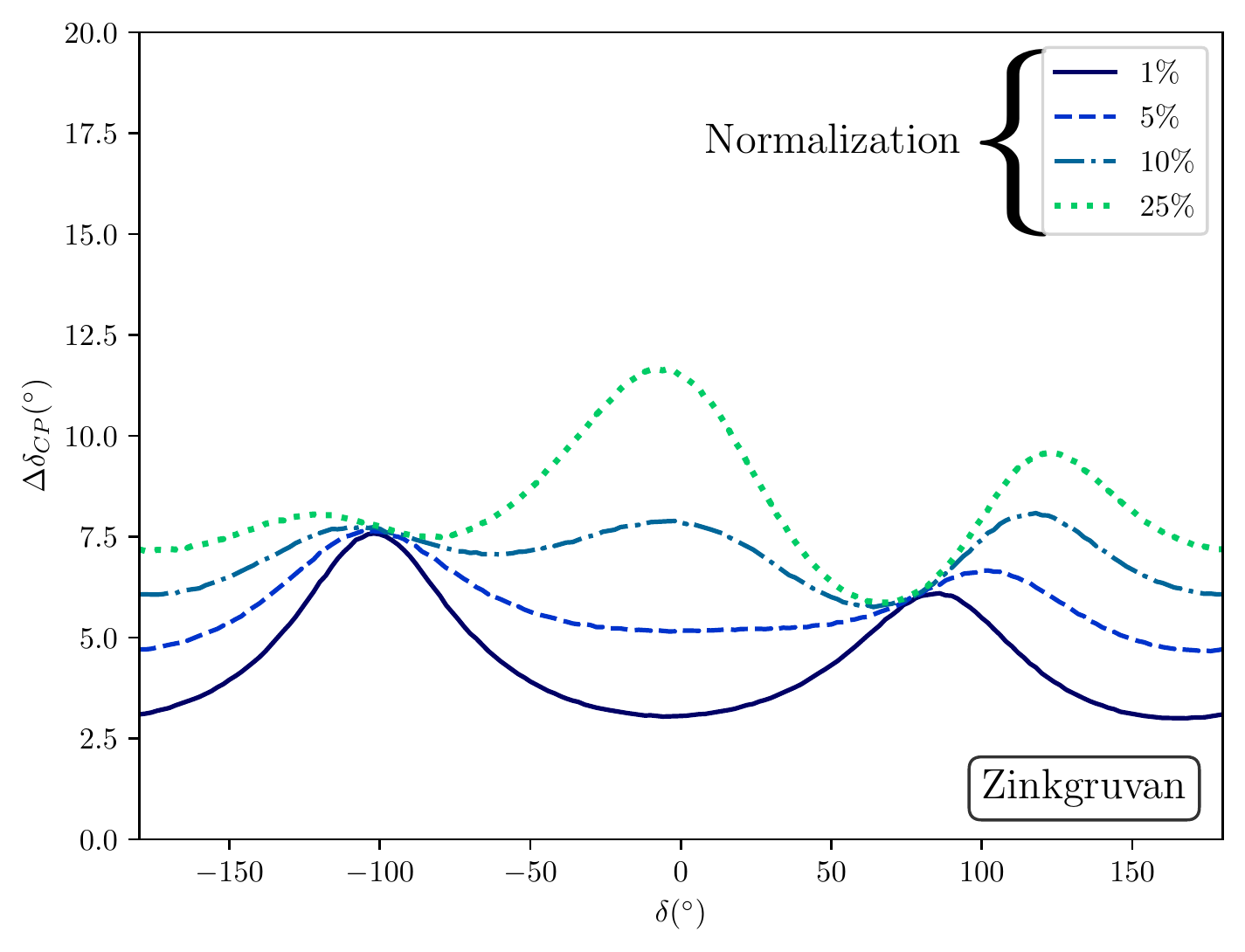}
\includegraphics[width=7.5cm]{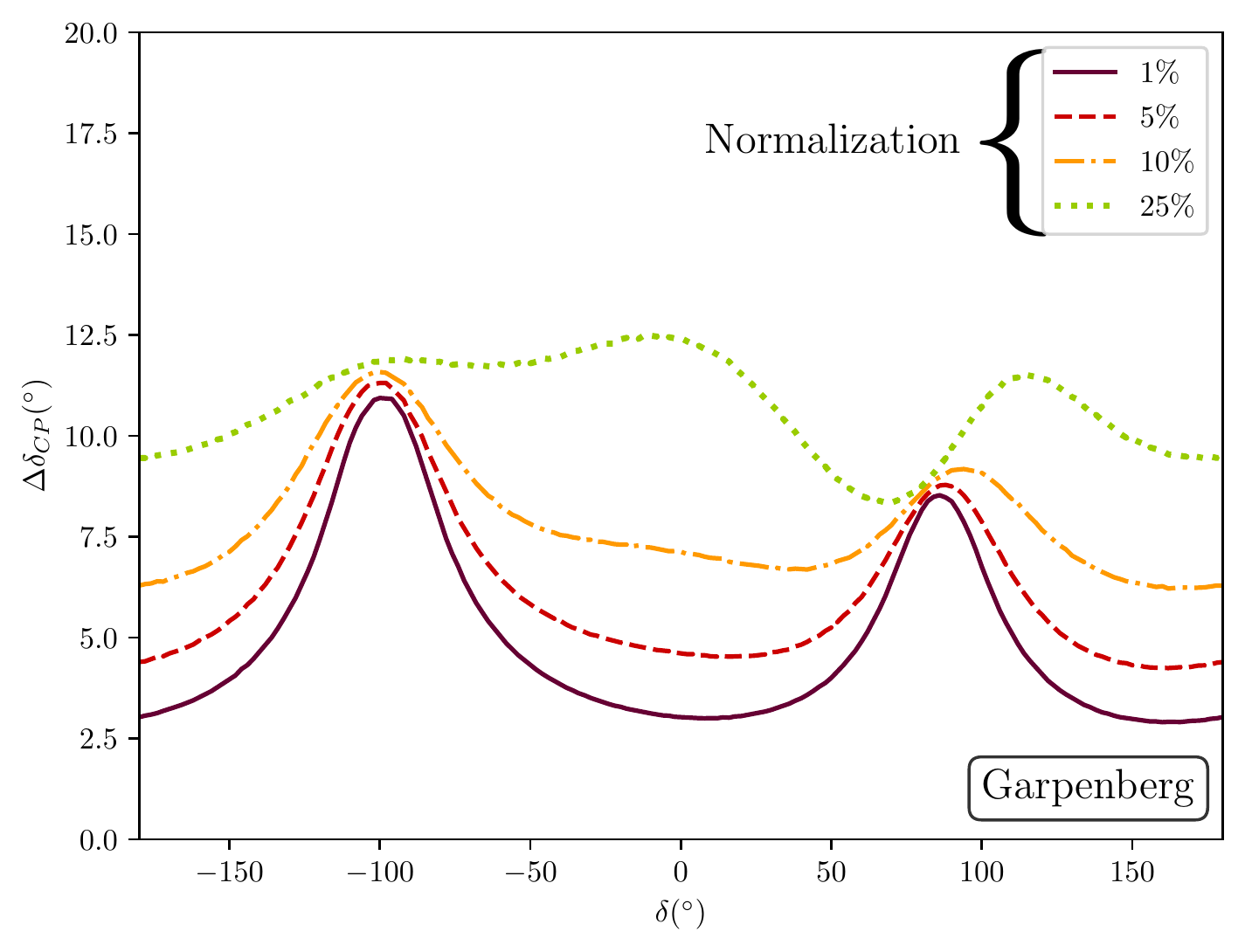}
\includegraphics[width=7.5cm]{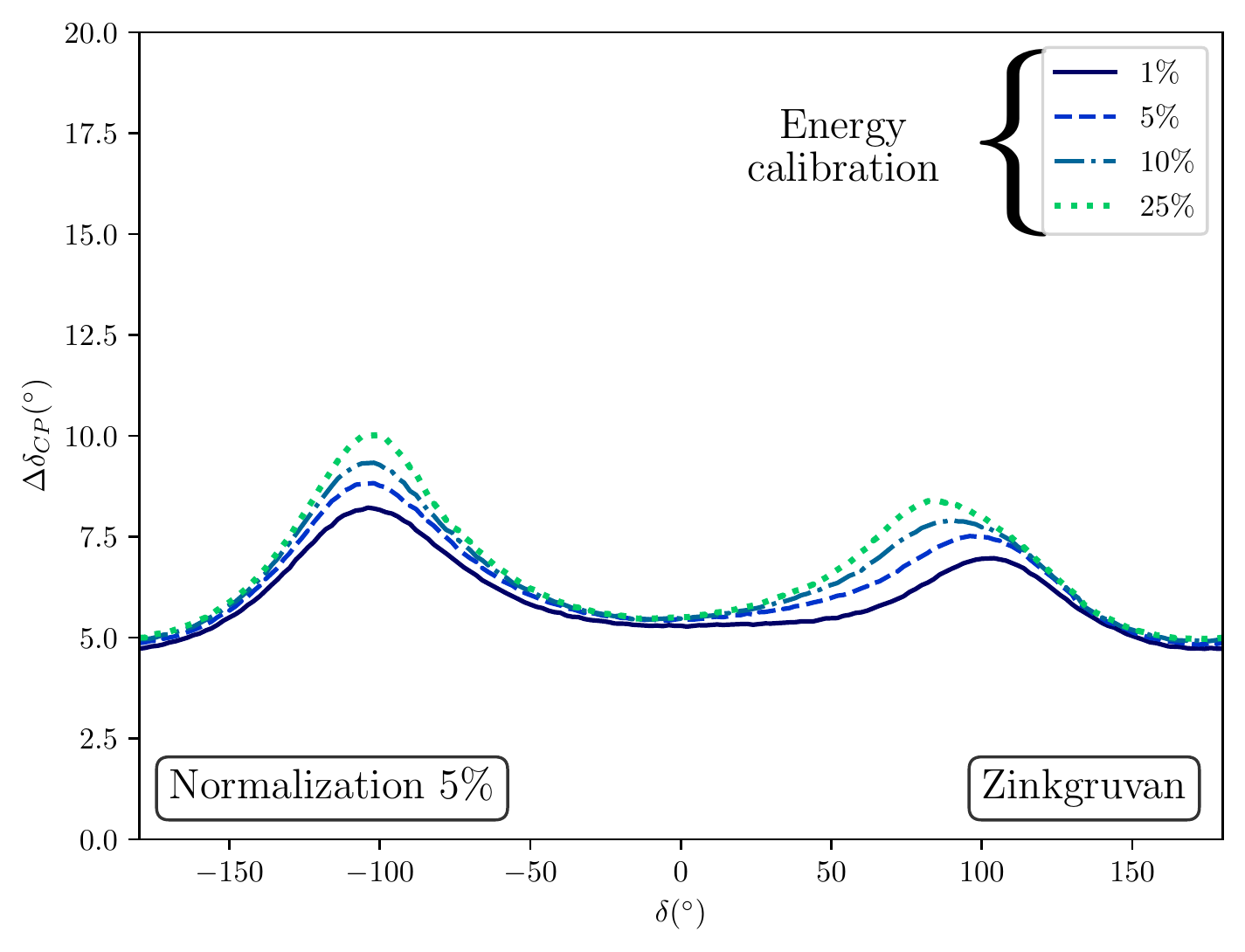}
\includegraphics[width=7.5cm]{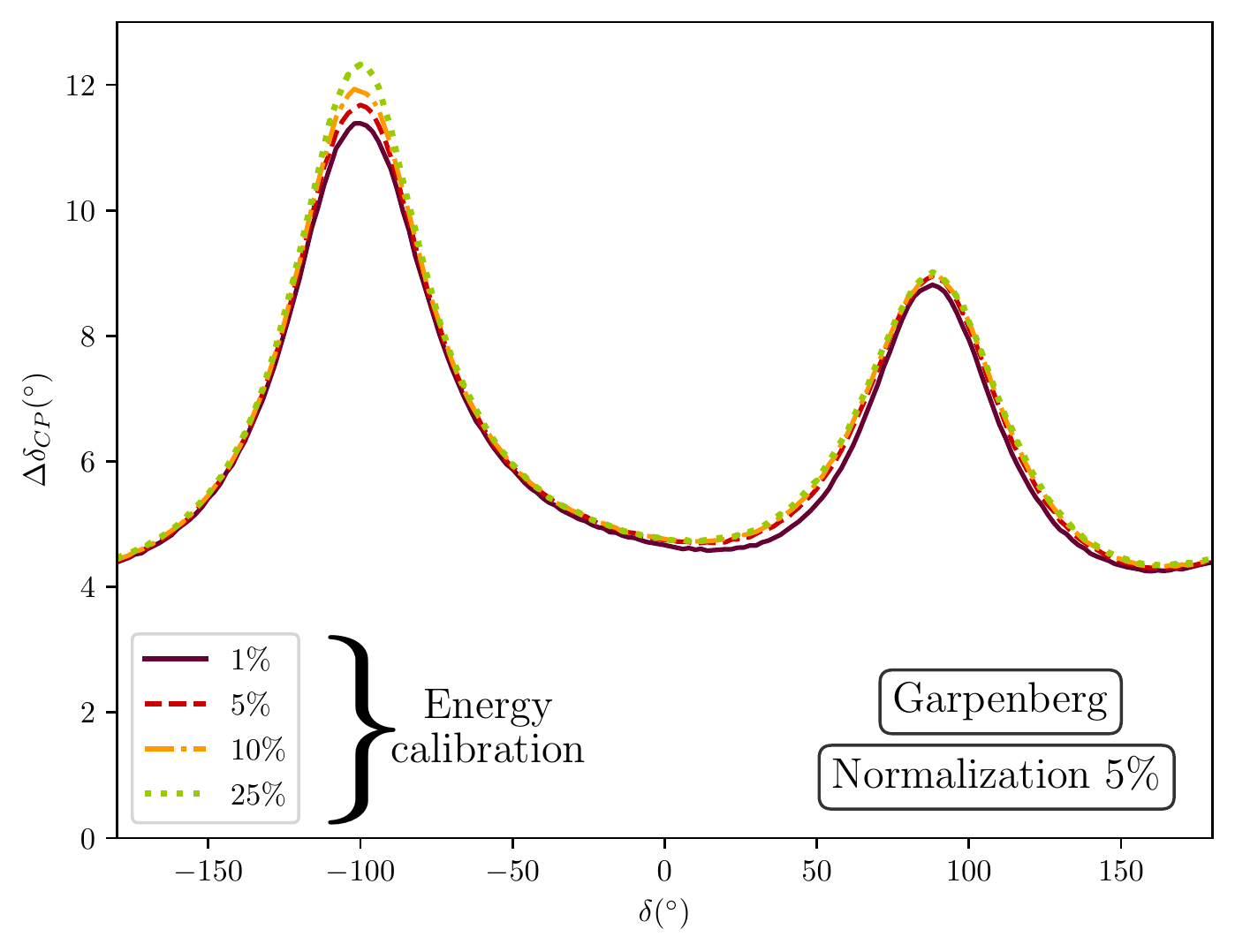}
\includegraphics[width=7.5cm]{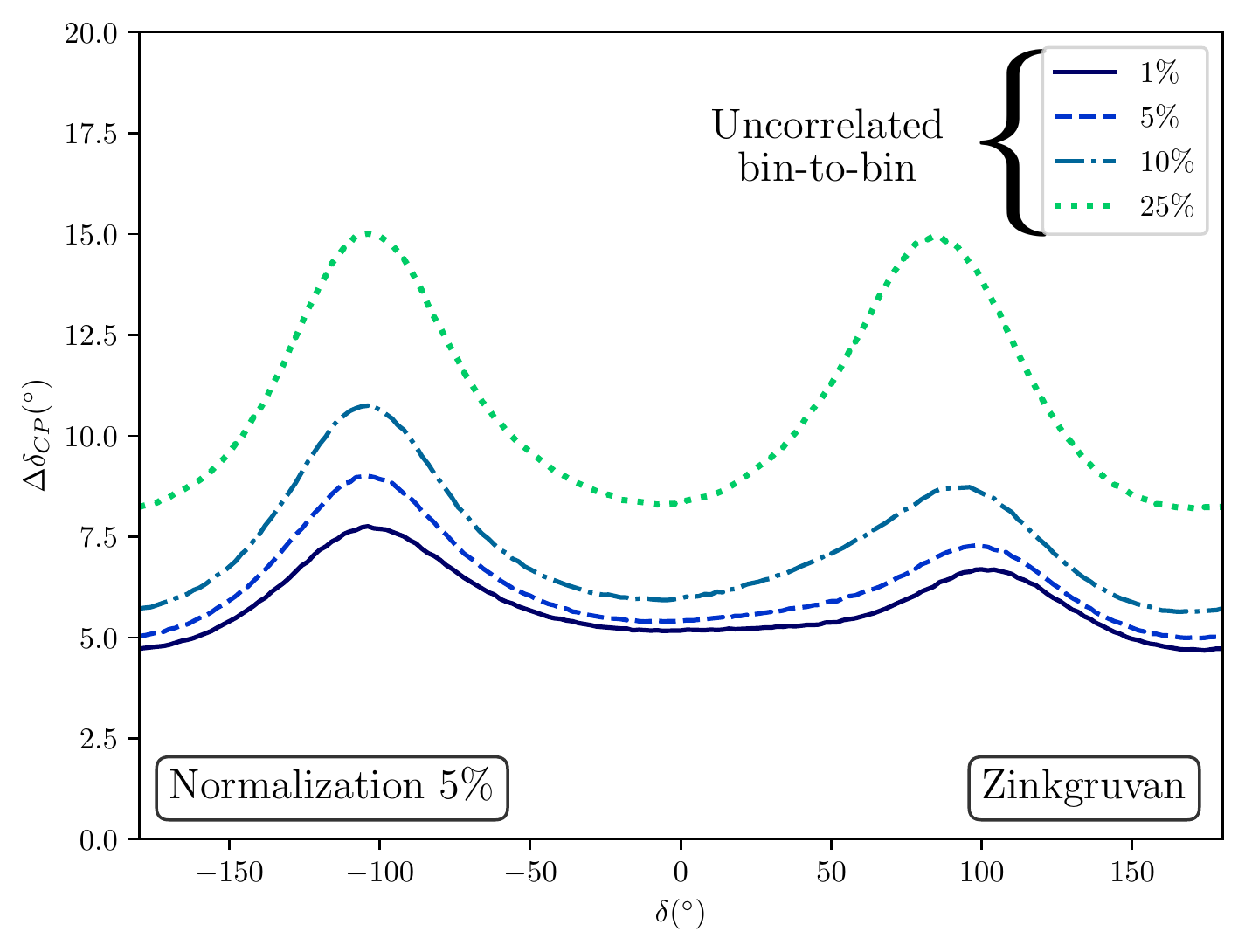}
\includegraphics[width=7.3cm]{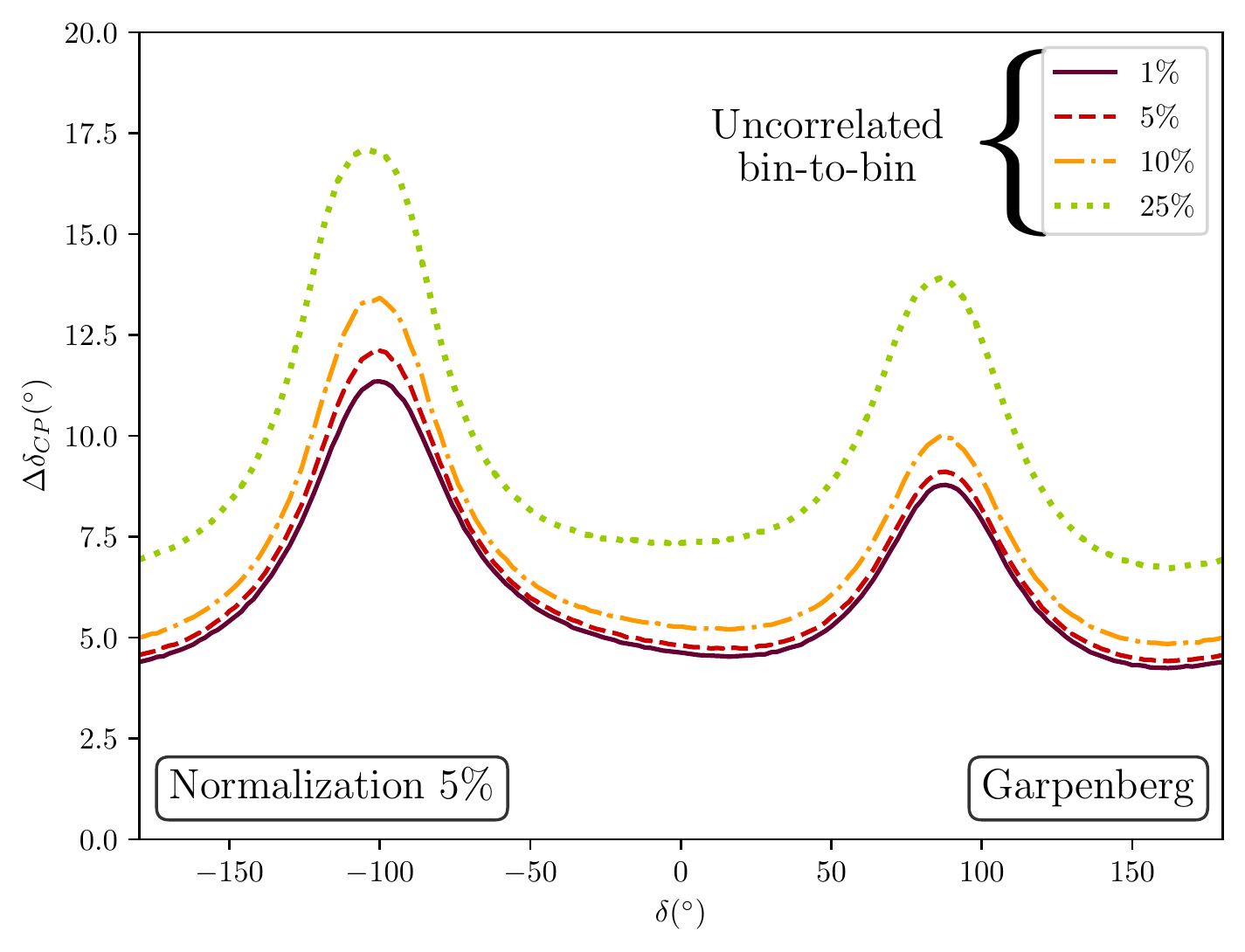}
\caption{Precision on the ESS$\nu$SB measurement of $\delta_{CP}$. Left (right) panels for the Zinkgruvan \SI{360}{\kilo\meter} (Garpenberg \SI{540}{\kilo\meter}) baseline.}
\label{fig:prec}
\end{figure}

In Figure~\ref{fig:prec}, we estimate the precision with which the ESS$\nu$SB will be able to measure the CP-violating phase $\delta_{CP}$. For each given possible value of $\delta_{CP}$, Figure~\ref{fig:prec} shows the standard error with which $\delta_{CP}$ is determined marginalizing over all nuisance parameters and oscillation parameters other than $\delta_{CP}$. We again observe a similar behaviour to that of Figure~\ref{fig:cpsens}. In particular, we find that the energy calibration uncertainty has a very minor impact, while the other two studied sources of uncertainties have a more important effect. Interestingly, the uncertainty on the overall normalisation is most important for values of $\delta_{CP} \sim 0$. Conversely, the bin-to-bin uncorrelated systematics that can also affect the shape of the recovered spectrum are more relevant close to maximally CP violating values, that is $\delta_{CP} \sim \pm \pi/2$.

\begin{figure}[ht]
\centering
\includegraphics[width=7.5cm]{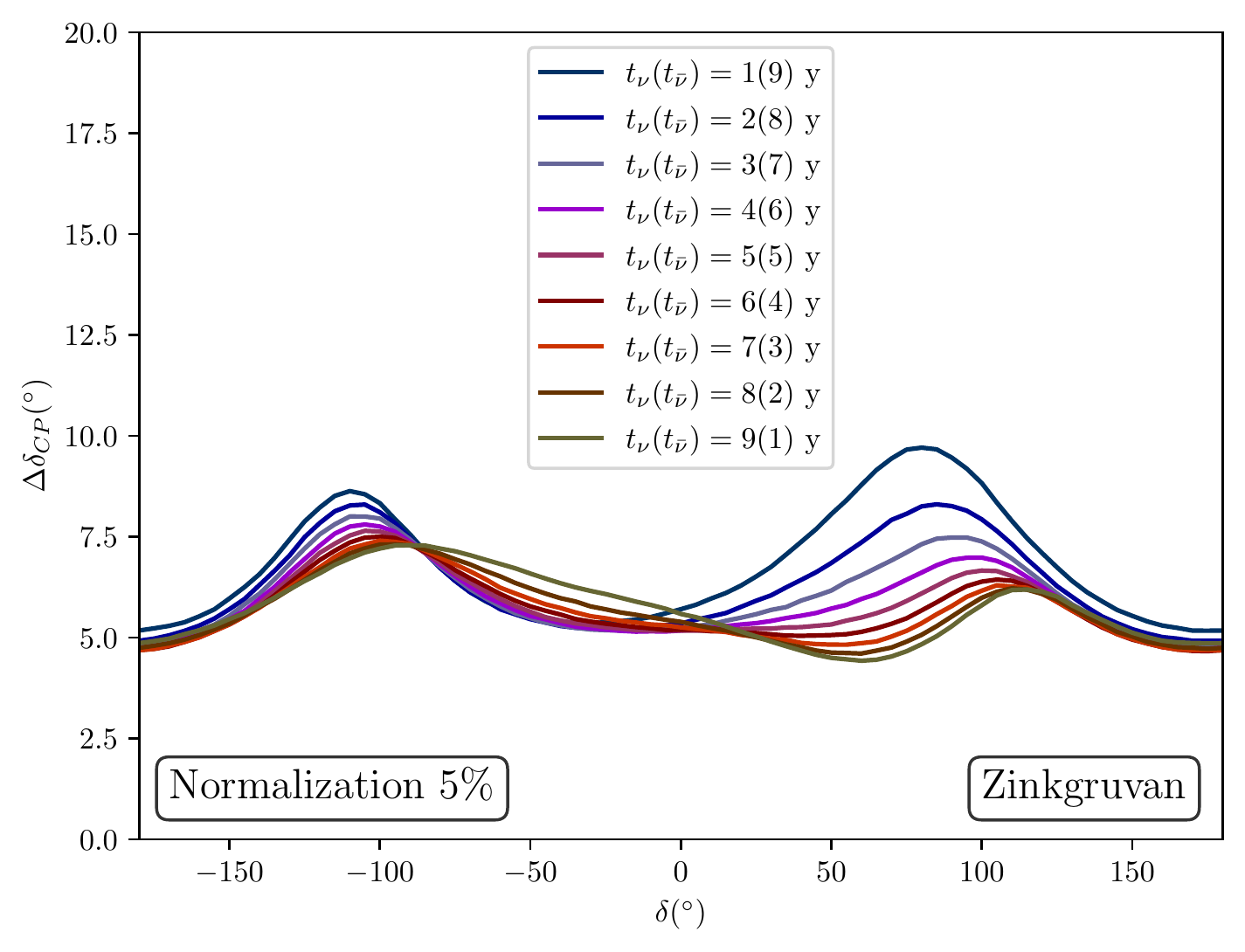}
\includegraphics[width=7.5cm]{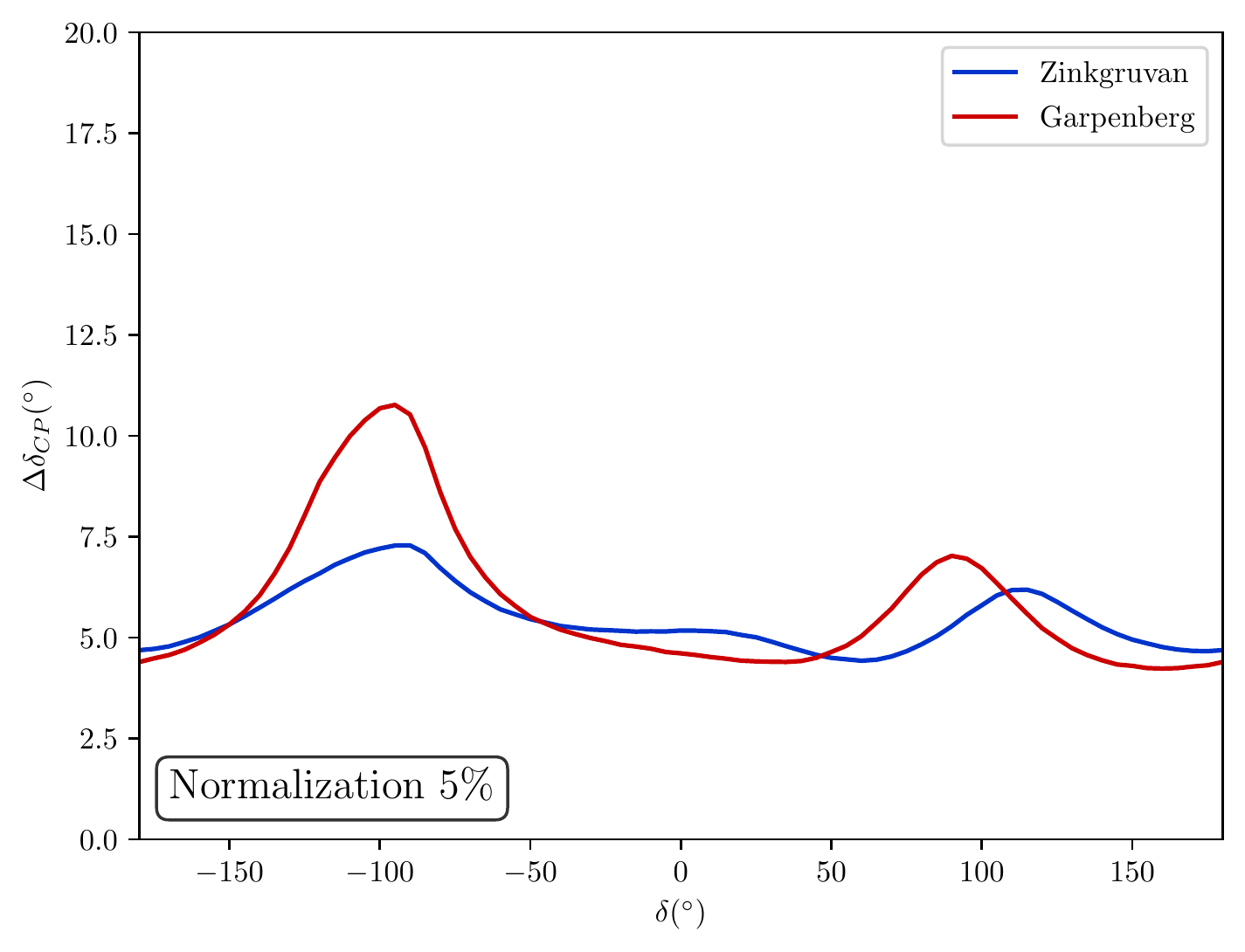}
\caption{Left panel, precision on the ESS$\nu$SB measurement of $\delta_{CP}$ for different splittings of the running time between neutrino and antineutrino modes at the Zinkgruvan baseline. The lines span from 9 (1) years to 1 (9) in (anti)neutrino, the total running time is always 10 years. Right panel, the precision on the measurement of $\delta_{CP}$ when the running time is optimised as in the left panel comparing the Zinkgruvan and Garpenberg options.}
\label{fig:optprec}
\end{figure}

Finally, in the left panel of Figure~\ref{fig:optprec} the dependence of the precision with which $\delta_{CP}$ would be measured is studied as a function of the splitting of the total running time between positive focusing (neutrino mode) and negative focusing (antineutrino mode). As an example, the Zinkgruvan (the \SI{360}{\kilo\meter}) option is shown, but the behaviour is very similar for Garpenberg (the \SI{540}{\kilo\meter}). As can be seen, the optimal splitting depends in the actual value of $\delta_{CP}$. Given the larger fluxes and cross sections, it is easier to accumulate statistics in neutrino mode and thus the best precision would be obtained by devoting longer periods of data taking to positive focusing. Conversely, around $\delta_{CP}=0$ or $\pi$ the complementary between the neutrino and antineutrino samples pays off and more even splits of the running time provide better sensitivity. The measurement strategy of the ESS$\nu$SB can profit from previous hints by preceding oscillation experiments and adapt the splitting between neutrino and antineutrino modes according to the left panel of Figure~\ref{fig:optprec}, depending on the value of $\delta_{CP}$ that the data point to. Following such a strategy, if the best splitting between neutrino and antineutrino modes is adopted for each value of $\delta_{CP}$, the precision that could be obtained is presented in the right panel of Figure~\ref{fig:optprec}. While around CP conserving values the precision achievable in the measurement of $\delta_{CP}$ is around $5^\circ$ for both the Garpenberg and Zinkgruvan options, Zinkgruvan outperforms Garpenbenberg around $\delta_{CP} = \pm \pi/2$, with the former providing a sensitivity better than $7^\circ$ for any possible value of $\delta_{CP}$. The conclusion of this study is thus that the ESS$\nu$SB experiment with its far detector located at \SI{360}{\kilo\meter}, in the Zinkgruvan site, can provide  unprecedented precision on the measurement of $\delta_{CP}$ ranging between $5^\circ$ and $7^\circ$ depending on its value. The same setup could deliver a $5 \sigma$ discovery of CP violation for a $71 \%$ of all possible values of $\delta_{CP}$.

\section{\textbf{Infrastructure and Safety}}
\label{InfraSafety}

\subsection{\textbf{Infrastructure and Conventional Facilities}}
The proposed upgrade of the European Spallation Source (ESS) to accommodate also the ESS neutrino Super Beam (ESS$\nu$SB), implies that the laboratory will have to host users of both the neutron spallation facility and of the neutrino beam facility. The upgrade would require several new buildings and facilities to be added to those already existing on the ESS site, which will have to be compliant with the conventional, cryogenic, radiological or other ESS safety requirements as well as with the Swedish regulations on environmental and sustainability aspects. The experience from the ESS and other similar facilities indicates that the cost of the technical infrastructure, site facilities upgrades and the new additions amount to the order of $30\%$ of the total cost of the project.

\subsubsection{\textbf{Buildings $\&$ General Safety}}
The ESS$\nu$SB project relies on using the existing high power LINAC of the ESS to deliver a \SI{5}{\mega\watt} beam of $H^-$ ions to an accumulator ring while concurrently providing the \SI{5}{\mega\watt} beam of protons for the spallation neutron source. The required modifications in the structure of the LINAC tunnel are expected to be minimal. The modifications that need to be studied, designed and implemented will concern the connection to the transfer tunnel and from there to the accumulator ring. To satisfy the requirements of radiation protection, the accumulator and target buildings will be located underground with the axis of the target modules at around \SI{20}{\meter} under the ground level. A geological study together with licensing for excavation will need to be performed prior to producing the layout design and site landscaping plan. The building design, construction and shielding shall follow the ESS safety rules, which tend to be more demanding and restrictive than the applicable national regulations. Definition and shielding specification documents for ESS$\nu$SB need to derive from the regulations of the Supervised and Controlled Radiation Areas and of ESS Radiation Protection Strategy for Employees that are currently valid for the existing ESS facility. Moreover, the target facility itself will need to have several buildings associated with it, for power, cooling, waste management, hot cells, target cavern and the decay tunnel at the downstream of the target.

\subsubsection{\textbf{Required Upgrade of the ESS Technical Utilities}}

Figure~\ref{fig:ESSBuild_Layout} shows a conceptual layout of the ESS site including the proposed modifications of the ESS$\nu$SB buildings. Several site/machine utilities will need to be added and/or upgraded to allow the ESS site to host the additional neutrino line, as summarised here below:

\begin{itemize}
\item The ESS$\nu$SB baseline will require installation of new cooling systems which are relying mainly on three coolants; air, water and helium saturated gas, and a heating, ventilation, and air conditioning (HVAC) system that is independent from the ESS one.

\item Delivering twice the power from the LINAC will require somewhat more than doubling the input power (the efficiency of the two beams will not be the same due to heavy chopping of the $H^-$ beam). 
This calls for an upgrade of the entire electrical (and grounding) network from the main High Voltage (HV) power station at the interface between ESS and the grid, the internal HV network from the HV station to the distribution sub-stations at the klystron gallery, the ancillary buildings of the ring and those of the target.

\begin{figure}[H]
  \centering
\includegraphics[width=1.\textwidth]{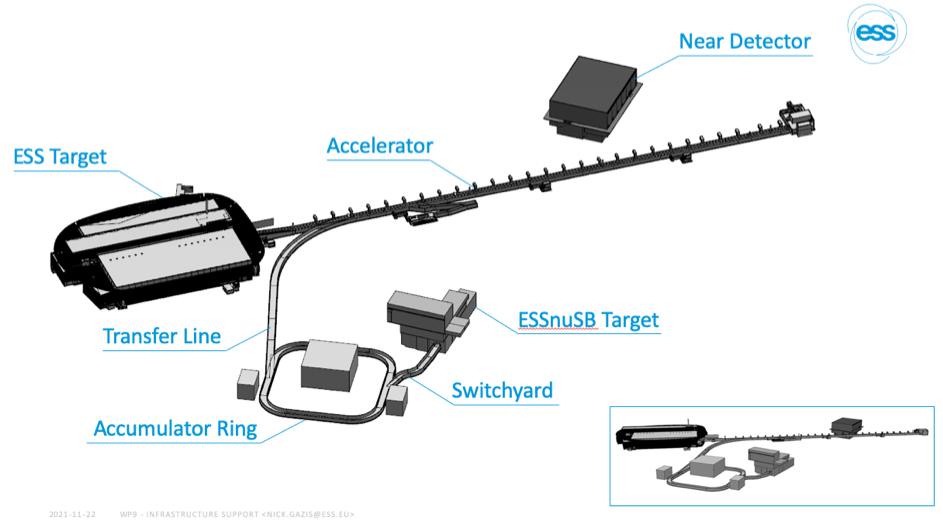}
  \caption{ESS building layout including the ESS$\nu$SB buildings.}
    \label{fig:ESSBuild_Layout}
\end{figure}

\item The transportation and handling of magnets and equipment to be installed at the new tunnels and accumulator ring is planned to be expedited via overhead cranes and wheeled crane bridges.
\item Structural requirement will need to be considered when sizing the cross-section of the new tunnels for ESS$\nu$SB. In Figure~\ref{fig:ess-upgrade}, the concept of the new ESS$\nu$SB buildings are highlighted in colour. Their dimensions follow a rough estimation and need to be further developed to match the ESSnuSB machine designs.
\end{itemize}

\subsubsection {\textbf{Far Detector Site}}

Two operating mines in Sweden, Zinkgruvan and Garpenberg, located \SI{360}{\kilo\meter} and \SI{540}{\kilo\meter} from the ESS site, respectively, have been selected as possible far detector sites for the ESS$\nu$SB project. The far detector infrastructure will be located at an average depth of $\sim$ \SI{1000}{\meter} below the ground surface. Measuring the rock strength and ground stress at the location of the neutrino detectors will be important for the design, construction, operation and safety of the infrastructure.

\subsection {\textbf{Safety}}
 
The ESS is a complex facility that will present several potential radioactive, as well as non-radioactive, hazards. Although ESS is not defined as a nuclear facility according to the Swedish regulation, it emphasizes the objective of setting radiation shielding and safety as a main priority for all phases of the project from design, through construction and operation, to decommissioning. This will be no different with the ESS$\nu$SB project.\\

\subsubsection{\textbf{General safety}} 
\label{sub:general safety}

Hazards originating from cryogenics, high-voltage, electromagnetic fields, heavy equipment, working on high heights, transports etc., are examples of non-radiation hazards. In order to protect the ESS staff, the public and the environment, it is necessary to define specific General Safety Objectives (GSO) according to ESS regulations. The GSO will serve as a guiding document, giving necessary input on how to design the ESS$\nu$SB facility. The work already done for the ESS proton LINAC~\cite{Jacobsson:IPAC11-WEPC166} will have to be revisited for the ESS$\nu$SB upgrade. At the time when the LINAC upgrade for the ESS$\nu$SB project starts, many of the procedures and internal safety regulations at ESS will be already well established. However, the safety aspects of the power upgrade of the linac from \SI{5}{\mega\watt} to \SI{10}{\mega\watt} will have to be assessed in view of inter alia the increased neutron radiation resulting from increased beam losses due to the stripping of H$^{-}$ ions and increased X-ray radiation due to the increase in the duty cycle of the accelerating cavities. 

\subsubsection{\textbf{Radiation safety}} 
\label{sub:radiation safety}

During operation, penetrating fast neutrons are generated in the target and by proton beam losses in the accelerator, leading to the production of radioactive nuclides, in particular in the target. Moreover, due to the high intensity of the primary proton beam delivered on the four targets, high interaction rates are also to be expected between the produced pions and muons and the different parts of the target station. A main objective will be \textit{to evaluate the radiation protection risks in the target station complex, in order to respect/obey the applicable ESS Radiation Safety Functions (RSF) (the IAEA standards + ESS specificities) to prevent or mitigate the radiological hazards, i.e. the dose uptake to on-site personnel and to the public}. In particular: the prompt and residual radioactive dose-equivalent rates, the radio-activation and radioisotopes formation, neutron flux and energy deposition estimation are the most important radiation safety parameters that will have to be studied in detail for the ESS$\nu$SB target station complex.

\subsubsection{\textbf{Shielding}}
\label{sub:Shielding}

The design of the target station shielding was optimised following the general radiation protection guidelines and were based on the prompt and residual dose calculations and the radio-activation analysis of the different parts of the target station facility. To provide a radiological shielding for the underground site surrounding the horn/target area, the helium vessel of the 4-horn gallery will be surrounded on all sides by a $\sim$~\SI{2.2}{\meter} thick iron inner-shield plus a 2~$\times \sim$~\SI{3}{\meter} thick concrete outer-shield. The preliminary simulations found also that a $\sim$~\SI{2.5}{\meter} thick concrete block must be placed up-stream to the baffle/collimator system, in order to protect the switch-yard area from the back scattered radiation from the horns and the targets. In order to protect the underground site down-stream of and around the decay tunnel and target-station beam-dump, these parts will be surrounded by a $\sim$~\SI{5.5}{\meter} thick concrete shield from all sides and with $\sim$~\SI{3.4}{\meter} alternating layers of iron blocks and a final concrete block down-stream to the beam dump.\\

In the residual dose and activation simulations, different cooling times: \SI{1}{\second}, \SI{1}{\hour}, 1 day, 10 days, 1- and 3-Months after switching off the beam, were assumed. The radiation-categorization of the specific zones of the target station complex was based on the Radiation Protection Ordinance (RPO) of the ESS and provided the following results; a) The 4-horn gallery, the decay tunnel and the beam dump areas are classified as \textit{prohibited-zones}, with no access permitted during operation. b) The power supply area and target station utility rooms are classified as \textit{controlled-zones}, with access permitted to authorized personal only. This is the work area above the decay tunnel and beam dump concrete shields, in which the annual radiation doses may exceed 3/10ths of the annual maximum permissible doses for exposed workers, i.e. $\sim$~3 $\mu$Sv/h. c) The control and data acquisition rooms located at the ground level that have a low dose rate of $<$ 1.5 $\mu$Sv/h, are classified as \textit{supervised-zone}, with access permitted all times.\\ 

\begin{figure}
    \centering
    \includegraphics[width=0.73\linewidth]{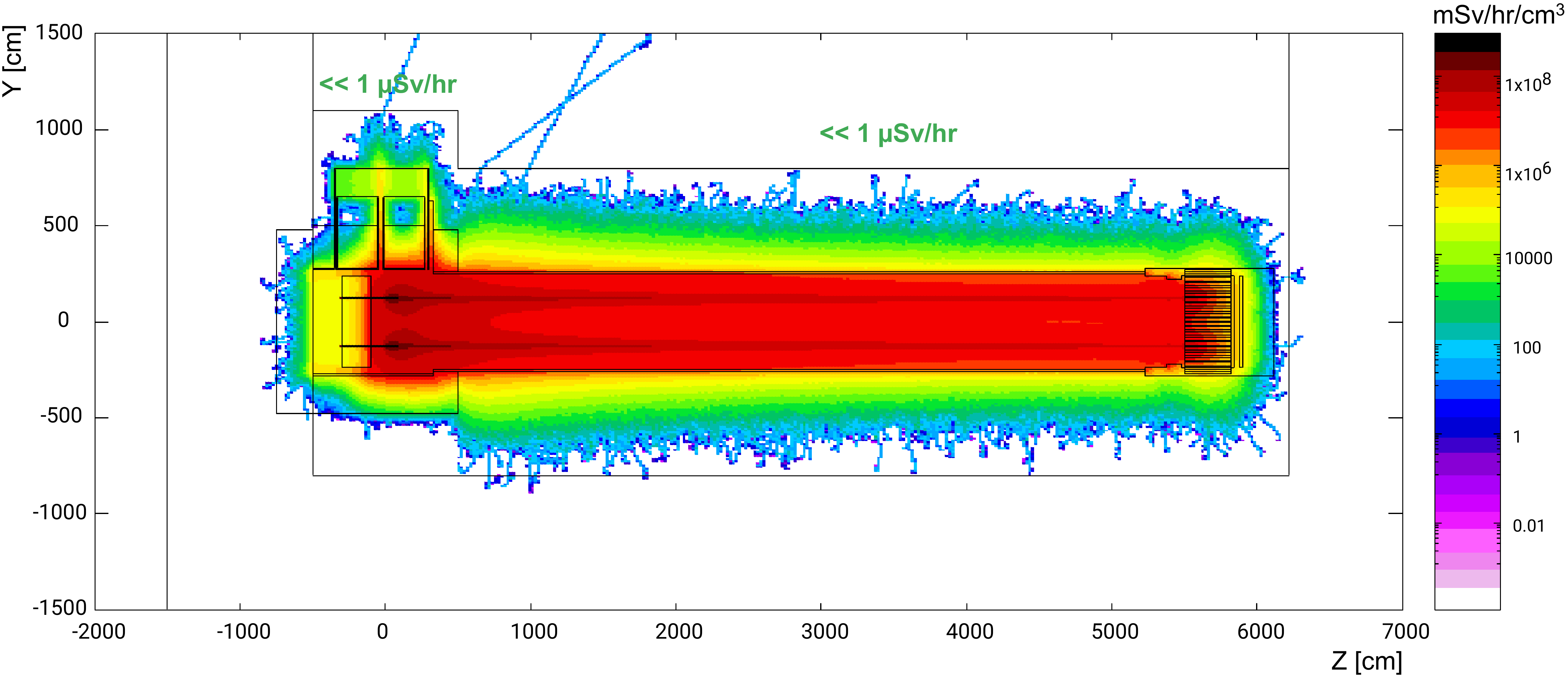}
    \includegraphics[width=0.73\linewidth]{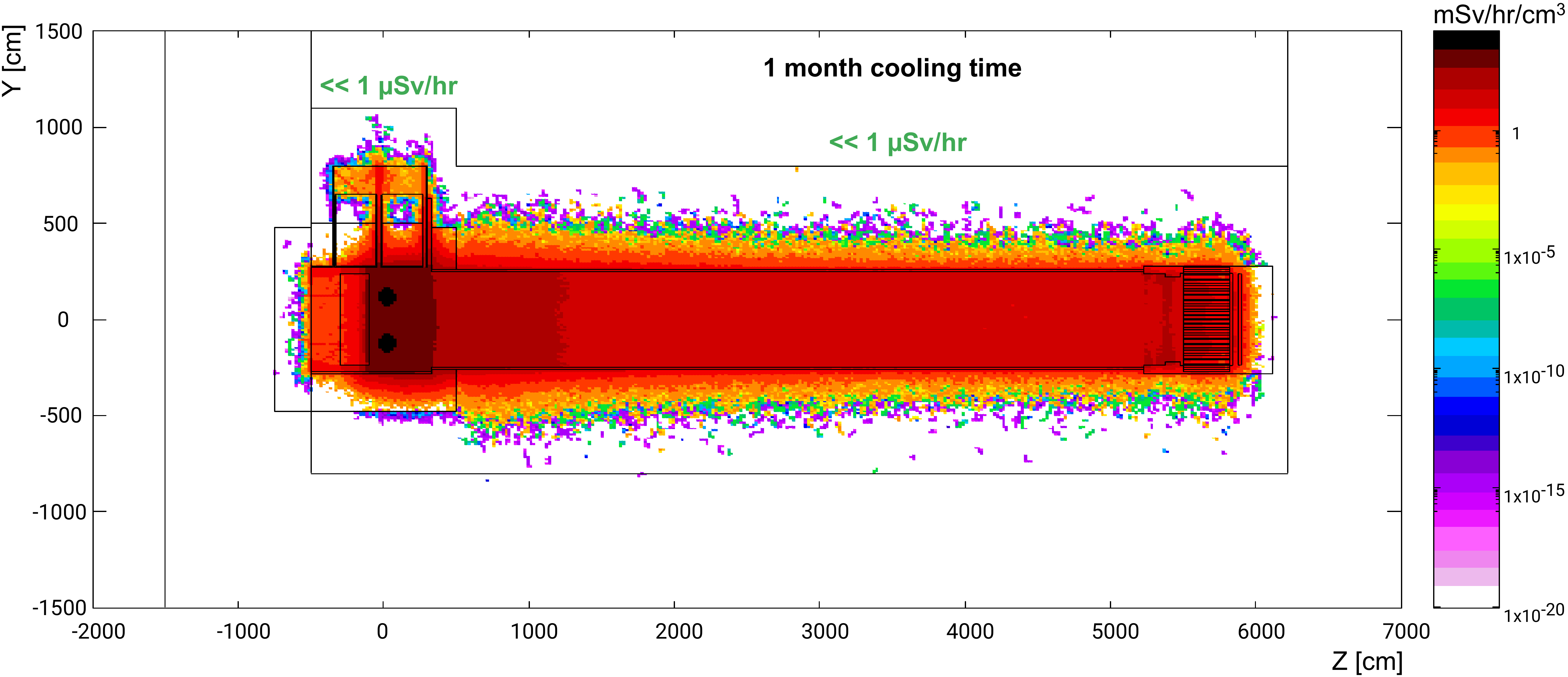}
    \caption{A radial view of the prompt (top) and residual, after one month cooling, (bottom) dose-eq rate distributions (in [mSv/h]) in the target station facility.}
    \label{fig:dose-eq}
\end{figure}

Figure~$\ref{fig:dose-eq}$ shows the prompt (top) and the residual, after one month cooling time, (bottom) dose rate distributions in the target station facility. The simulations show, in the top frame, that during irradiation the dose-equivalent rate reaches an acceptable level of less than 1 $\mu$Sv/h at the ground level above the 4-horn system, the decay tunnel, the beam dump, and the underground site/earth behind the beam dump, indicating the efficiency of the shielding system and allow for these zones to be accessible during irradiation. The bottom frame shows that the residual dose-equivalent values, after one month cooling time, for all designed accessible areas are also below 1 $\mu$Sv/h. The residual dose-equivalent rates, the radio-activation, the neutron flux, the produced radio-isotope and the deposited energy distributions were studied, for each part of the target station complex, as a function of the cooling time (from \SI{1}{\second} to 3 months).\\

The stopping power of the proposed beam dump with respect to the unabsorbed primary protons and the secondary $\mu^{\pm}$ and $\pi^{\pm}$ particles has also been studied.. The particle yield, $\eta_{par}$ ($\it{par}$ is particle type), was measured at different boundaries of the beam dump: \textbf{a}) at the upstream face of the BD, \textbf{b}) at the down-stream face of the BD core and \textbf{c}) at the lower-stream face of the last concrete block of the beam dump structure. The core stopping power is defined as the ratios of $\eta_{par}$ at \textbf{b} and \textbf{c} to that at \textbf{a}. The stopping power at \textbf{b}, is found to be $>$ 98$\%$ for all scored particles. Moreover, no particles were found exiting the boundary surface \textbf{c}, i.e. no particles passing through the beam dump to the underground site behind it, with respect to the number of generated pot. This indicates that the overall protection performance complies with the safety requirements of the ESS underground site.

\section{ESS$\nu$SB Synergies with Future Projects}

\subsection{\textbf{Low Energy nuSTORM}}
\label{lenustorm}

The basic idea of nuSTORM is to store muons, generated from pion decays, in a racetrack storage ring and use the muon- and electron-neutrinos that are created, from the muon decays, in one of the two straight sections to form a beam that can be used to measure neutrino cross-sections as well as search for sterile neutrinos. Contrary to a neutrino beam generated from pion decay, which contains nearly muon neutrinos only, like that of the ESS$\nu$SB, the nuSTORM neutrino beam will contain equal amounts of muon- and electron-neutrinos, thus making high statistics measurements of both electron- and muon-neutrino cross-sections possible. In particular, precise electron-neutrino nuclear cross-sections are needed for the interpretation of the electron-neutrino spectrum detected by the ESS$\nu$SB Far Detector. So far, nuSTORM design studies have been made for the Fermilab and CERN accelerators with proton energies in the order of \SI{100}{\giga\electronvolt} \cite{adey2013nustorm,Ahdida:2654649}. Pions produced from protons of such energies have an average energy of ca \SI{5}{\giga\electronvolt} and will decay to muons with an average ca \SI{4}{\giga\electronvolt} energy, which will in turn decay to neutrinos of average ca \SI{3}{\giga\electronvolt} energy. The length of the racetrack straight sections was chosen to be about \SI{180}{\meter}. Even if the neutrino momentum distribution so produced is broad, it will hardly cover the neutrino momentum distribution of ESS$\nu$SB, which has an average energy of ca \SI{0.4}{\giga\electronvolt}, nor the neutrino momentum distribution of the Hyper-K, for which the average energy is ca \SI{0.6}{\giga\electronvolt}.\\

\begin{figure}[hbt]
  \begin{center}
    \includegraphics[width=0.65\linewidth]{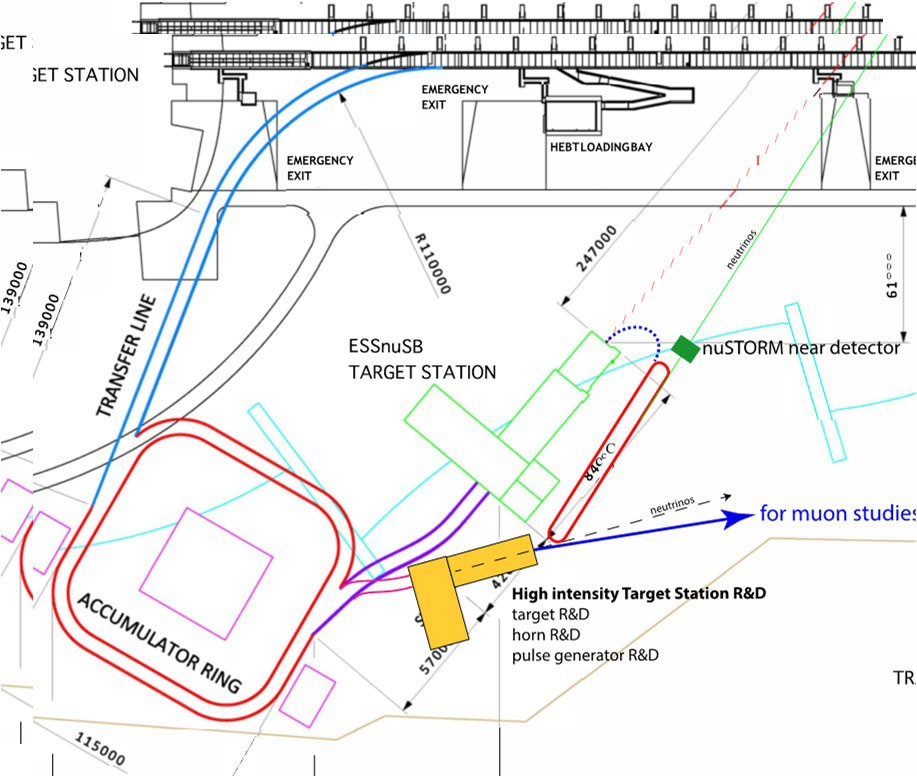}
\caption{Layout of the ESS$\nu$SB with the proposed low energy nuSTORM racetract ring (in red) and a dedicated target station (in yellow). (Units in mm)}
    \label{fig:nuSTORM}
  \end{center}
\end{figure}

The ESS-based Low Energy nuSTORM (LEnuSTORM) would use protons of \SI{2.5}{\giga\electronvolt} from which a neutrino average energy of ca \SI{0.4}{\giga\electronvolt} will be obtained. It would be difficult to generate a sufficiently powerful beam from the CERN PS, \SI{1.4}{\giga\electronvolt}, Booster or the CERN, \SI{26}{\giga\electronvolt}, PS to cover the low neutrino energies of ESS$\nu$SB and Hyper-K with sufficient statistics. In view of this, it is proposed to carry out a design study of an LEnuSTORM using the ESS$\nu$SB-upgraded ESS LINAC. The target to produce the muons could be either the already designed targets used for producing the neutrino Super Beam with the \SI{1.3}{\micro\second} proton pulses or a new dedicated target. The length of the racetrack straight sections will be much shorter than in the Fermilab and CERN designs and the required strength of the large aperture magnets in the racetrack arcs will be much lower. The lattice design could be of the Fixed-Field Alternating-Gradient (FFAG) accelerator or FODO type or a mixture of both. Figure~\ref{fig:nuSTORM} shows a lay-out of ESS$\nu$SB on the ESS site with proposed positions of the LEnuSTORM racetrack ring and, in this case, a dedicated target station. The LEnuSTORM straight section is directed such that the neutrino beam produced will first hit a LEnuSTORM Near Detector indicated in this lay-out and then the ESS$\nu$SB Near Detector, which is not visible in this figure but located to the right and just above the figure, that would be used as the far detector for the LEnuSTORM beam.

\subsection{\textbf{A Proton Complex Test Facility for a Muon Collider at ESS}}
\label{muoncollider}

It is proposed to make, as part of the International Muon Collider design study project \cite{muoncolider}, a design study of a Muon Collider Proton Complex Test Facility which will be based on the use of the ESS LINAC, on the already designed ESS$\nu$SB accumulator ring and on a new compressor/buncher ring. If CERN should decide around 2030 to go forward with the construction of a high energy Muon Collider with a collision energy of 3, 10 or \SI{14}{\tera\electronvolt}, the construction of such a Proton Complex Test Facility at ESS could be started around the same time for the purpose of demonstrating that \SI{2}{\nano\second} proton pulses of 10$^{14}$-10$^{15}$ protons at a rate of \SI{14}{\hertz} can actually be produced in practice. Such a test facility at ESS would require substantial funding but even more funding would be required to start the construction at CERN of such a test  facility, as the build-up of a \SI{5}{\mega\watt} proton accelerator would have to be initiated and this well before 2030 to be in time. Moreover, with such a Proton Complex in operation at ESS and with muon cooling at low intensity being demonstrated in practice in a test facility at CERN, this would open the way - in a longer perspective - for the construction at ESS of a \SI{125}{\giga\electronvolt} Higgs Factory Muon Collider with a unique potential for measurements of the Higgs self-coupling, extremely rare decays and the width of the Higgs boson \cite{rubbia2019searches} and a 3, 10 or \SI{15}{\tera\electronvolt} Muon Collider at CERN for Energy Frontier experiments.\\

\begin{figure}[hbt]
  \begin{center}
    \includegraphics[width=0.55\linewidth]{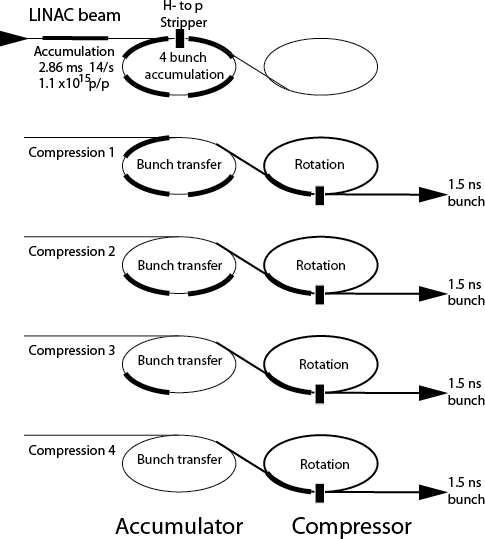}
\caption{The proposed accumulator ring layout dedicated for the Muon Collider Complex at ESS.}
    \label{fig:accumulator_LEnuSTORM}
  \end{center}
\end{figure}

The design study of a Muon Collider Proton Complex at ESS would be based on, inter alia, a faster chopping scheme for the LINAC, a new operation scheme for the accumulator ring, a new design of a compressor/bunch rotation ring and, in a second phase, a separate target station with a target and capture system (horn or solenoid) that could withstand the \SI{2}{\nano\second} short bunches of 10$^{15}$ protons. The basic principle for the generation of the \SI{2}{\nano\second} long pulses from the \SI{2.86}{\milli\second} 10$^{15}$ proton LINAC pulses is illustrated in Figure~\ref{fig:accumulator_LEnuSTORM}. The LINAC pulse is chopped into many short pulses that are injected into the accumulator ring and then extracted into the compressor/buncher ring where they are phase-rotated to ca \SI{2}{\nano\second} length (\SI{1.5}{\nano\second} in the Figure). This calls for the development of a high frequency chopper acting at the level of the LINAC H$^{-}$ source and an adaptation of the accumulator ring acceptance, RF system, timing and optics. As to the design of the accumulator and the compressor/buncher rings, there has been a design based on the use of the \SI{5}{\giga\electronvolt} \SI{4}{\mega\watt} SPL proton LINAC, that was planned for construction at CERN \cite{protonRutherford} as well as a design based on the use of the \SI{8}{\giga\electronvolt} high power Project-X proton LINAC, that was planned at Fermilab \cite{Flanagan:2010zza}. These designs will be used as starting points for the design and simulation of a compressor/buncher ring based on the use of the ESS LINAC. In Figure~\ref{fig:nuSTORM}, there is an indication of the direction of the ejected \SI{2}{\nano\second} pulsed muon beam towards an area at ESS, where there is free space for a second phase project to use the beam so-produced to build and test a target station and cooling front-end set-up there. In Figure~\ref{fig:nuSTORM}, the compressor ring is tentatively assumed to be located in the same tunnel as the accumulator.

\section{Proposed Project Time Line} \label{timeplan}

The proposed time frame for accomplishing the objectives of the experiment, can be shown in the following order:
\begin{itemize}
  \item 2022: End of ESS$\nu$SB Conceptual Design Study, CDR, and preliminary costing
  \item 2023-2026: Civil engineering and LEnuSTORM Conceptual Design Study
  \item 2026-2027: TDR, Preparatory Phase 
  \item 2027-2030: Preconstruction Phase, International Agreement 
  \item 2030-2037: Construction of the neutrino beam facility and the detectors, including commissioning
\end{itemize}

According to the report \cite{ESSops}, the schedule for normal operations of the ESS neutron source as from 2030 and onward will include two long shutdowns per year:
\begin{itemize}
  \item Winter (ca 8.9 weeks)
  \item Summer (ca 8.1 weeks)
\end{itemize}

This amounts to a total of 17 weeks of shutdowns per year, which will be available for the work to upgrade the LINAC.

\section{Conclusion}

The discovery of a large - in comparison with assumptions made before 2012 - value of $\theta_{13}$ implies that the influence of irreducible systematic errors in the search for, and measurements of, leptonic CP violation is close to three times lower at the second neutrino oscillation maximum as compared to the first. As it is the systematic errors that currently limit the accuracy in the measurement of the CP violation using neutrino long-baseline experiments, measuring at the second oscillation maximum represents a crucial advantage. However,  making measurements at the 3 times more distant second maximum requires a significantly more intense neutrino beam to keep the statistical errors comparable to the systematic errors, implying the need for an exceptionally powerful proton driver.\\

The ESS$\nu$SB project proposes to use the world-uniquely powerful \SI{5}{\mega\watt} ESS proton LINAC to produce a very intense neutrino beam, and place the far detector at a distance corresponding to the second oscillation maximum. With the use of a \SI{500}{\kilo\tonne} fiducial mass Cherenkov detector, it has been demonstrated that ESS$\nu$SB will reach, after 10 years of data taking, more than 70\% $\delta_{CP}$ discovery coverage with a significance larger than 5~$\sigma$. After discovery of leptonic CP violation, ESS$\nu$SB will measure $\delta_{CP}$ with a a standard error smaller than 8$^\circ$ for all values of $\delta_{CP}$.\\

Moreover, ESS$\nu$SB has a high potential for future upgrades by using the muons produced at the same time as the neutrinos for the realisation of a future low energy nuSTROM and, in a longer term perspective, a Muon Collider.

\section{Acknowledgments}

This project has been supported by the COST Action EuroNuNet “Combining forces for a novel European facility for neutrino-antineutrino symmetry-violation discovery” and by the European Union’s Horizon 2020 research and innovation programme under grant agreement No 777419. We also acknowledge support of the funding agencies CNRS/IN2P3–France, the Deutsche Forschungsgemeinschaft (DFG, German Research Foundation)-Projektnummer 423761110, the Bulgarian National Science Fund Contract DCOST01/8, the Ministry of Science and Education of Republic of Croatia grant No. KK.01.1.1.01.0001 and the Spanish Agencia Estatal de Investigacion through the grants IFT Centro de Excelencia Severo Ochoa No CEX2020-001007-S and PID2019-108892RB funded by MCIN/AEI/10.13039/501100011033.

\clearpage
\clearpage

\bibliographystyle{elsarticle-num}
\bibliography{biblio}

\end{document}